\documentclass[twocolumn,trackchanges]{aastex61}

\usepackage{amsmath,amsthm,amssymb}

\received{}
\revised{}
\accepted{}
\submitjournal{ApJ}

\shorttitle{Numerical Simulations of Flare-productive Active Regions}
\shortauthors{Toriumi \& Takasao}

\begin{document}

\title{Numerical Simulations of Flare-productive Active Regions: $\delta$-sunspots, Sheared Polarity Inversion Lines, Energy Storage, and Predictions}

\correspondingauthor{Shin Toriumi}
\email{shin.toriumi@nao.ac.jp}

\author[0000-0002-1276-2403]{Shin Toriumi}
\affiliation{National Astronomical Observatory of Japan, 2-21-1 Osawa, Mitaka, Tokyo 181-8588, Japan}
\nocollaboration

\author{Shinsuke Takasao}
\affiliation{Department of Physics, Nagoya University, Furo-cho, Chikusa-ku, Nagoya, 464-8602 Aichi, Japan}
\nocollaboration



\begin{abstract}
Solar active regions (ARs)
that produce strong flares
and coronal mass ejections (CMEs)
are known to have a relatively high non-potentiality
and are characterized by $\delta$-sunspots
and sheared magnetic structures.
In this study,
we conduct a series of
flux emergence simulations
from the convection zone to the corona
and model four types of active regions
that have been observationally suggested
to cause strong flares,
namely the Spot-Spot, Spot-Satellite, Quadrupole, and Inter-AR cases.
As a result,
we confirm that $\delta$-spot formation
is due to the complex geometry and interaction
of emerging magnetic fields,
with finding that the strong-field, high-gradient, highly-sheared
polarity inversion line (PIL)
is created by the combined effect
of the advection, stretching, and compression
of magnetic fields.
We show that free magnetic energy builds up
in the form of a current sheet
above the PIL.
It is also revealed that
photospheric magnetic parameters
that predict flare eruptions
reflect the stored free energy
with high accuracy,
while CME-predicting parameters
indicate the magnetic relationship
between flaring zones and entire ARs.
\end{abstract}

\keywords{Sun: corona --- Sun: flares --- Sun: interior --- Sun: photosphere --- (Sun:) sunspots}



\section{Introduction}\label{sec:intro}

Strong flares
and coronal mass ejections (CMEs),
the most catastrophic energy-releasing events
in the solar system,
are known to occur
in active regions (ARs)
including sunspots
\citep{pri02,shi11}.
Numerous observations have revealed that
complex ARs called
``$\delta$-sunspots,''
in which the umbrae
of positive and negative polarities
share a common penumbra,
tend to produce
powerful flare eruptions
\citep{kue60}\footnote{The other classifications of the Mount Wilson magnetic classification are $\alpha$ (unipole), $\beta$ (bipole), and $\gamma$ (multiple spots with intermixed polarity).}.
According to statistical studies
by \citet{shi94}, \citet{sam00}, and \citet{guo14},
more than 80\% of {\it GOES} X-class flares
occur in $\delta$-spots.
In such ARs,
flare eruptions often occur
in sheared magnetic structures
above polarity inversion lines (PILs).
Many observers have pointed out the importance
of strong-field, high-gradient, highly-sheared PILs
\citep[e.g.,][]{hag84,tan91,zir93,fal02,sch07}.
Therefore,
for understanding flare eruptions,
it is essential to reveal
the formation of
$\delta$-spots and such sheared structures
and its relation with the evolution
of entire ARs
\citep[for further review see][]{wanh15}.

\begin{figure*}
  \begin{center}
    \includegraphics[width=150mm]{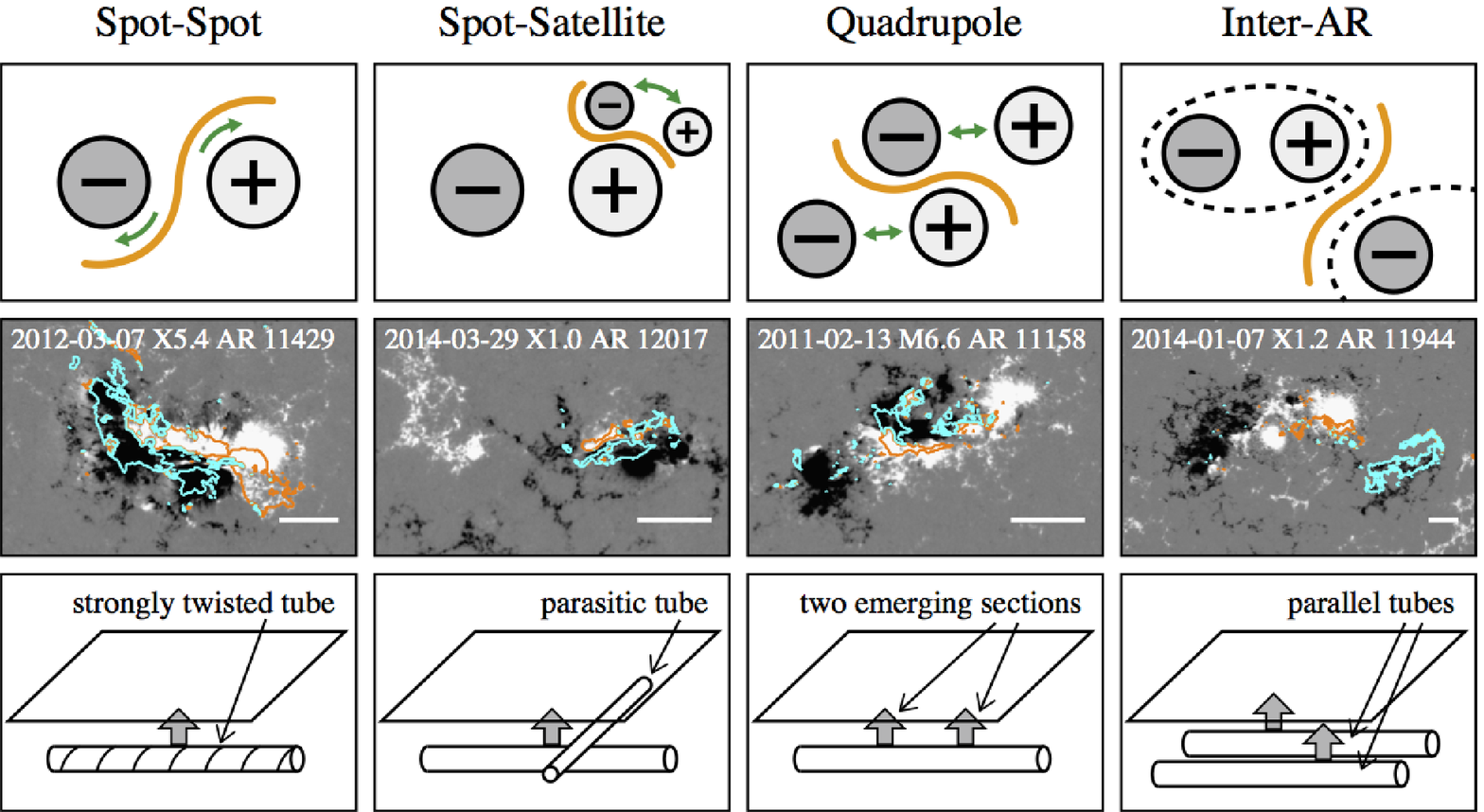}
  \end{center}
  \caption{Four categorizations
    of flaring ARs.
    Top row:
    polarity distributions,
    in which sunspots are indicated by circles
    with plus and minus signs.
    Sheared PIL involved in flare eruptions
    is shown with an orange line,
    whereas the proper spot motions
    are indicated with green arrows.
    Second row:
    sample flare events.
    {\it SDO}/HMI magnetogram
    is shown as background
    and the orange and turquoise contours
    indicate the flare ribbons
    detected by AIA 1600 {\AA}
    in the positive and negative polarities,
    respectively.
    Date, {\it GOES} flare class, and NOAA AR number
    are presented.
    White bar indicates
    a length of $50\arcsec$
    ($\sim 36.3\ {\rm Mm}$).
    Bottom row:
    schematic diagrams
    showing the numerical setup
    of the four simulation cases
    (Section \ref{sec:cases}).
    Top and second rows
    reproduced from \citet{tor17}.
    \label{fig:model}}
\end{figure*}

Recently,
\citet{tor17} surveyed all ARs
that produced $\geq$ M5.0-class events
in Solar Cycle 24
(events within 45$^{\circ}$
from the disk center
in 6 yr from May 2010)
and classified them
into four categories
based on pioneering work
by \citet{zir87}:
see also \citet{taki15}.
Figure \ref{fig:model} summarizes
the four categories.
\begin{description}
\item[Spot-Spot] A complex, compact $\delta$-spot group,
  in which a large, long sheared PIL
  extends across the whole AR.
  A representative region is NOAA AR 11429,
  which produced an X5.4-class event
  along the central PIL.
  \citet{tak15} suggested the possibility that
  this AR is created through the emergence
  of a strongly twisted, kink unstable flux tube
  \citep[see also][]{tan91,lin96,fan99}.
\item[Spot-Satellite] A newly emerging minor bipole
  appears in the close vicinity
  of one of the preexisting main sunspots
  and creates a small $\delta$-spot
  with a compact PIL
  between the main and satellite spots.
  NOAA 12017,
  producing an X1.0 event,
  falls into this category
  \citep{kle15}.
\item[Quadrupole] A $\delta$-configuration
  is formed by the collision
  of opposite polarities
  from two emerging bipoles
  of comparable size.
  A typical example
  is NOAA 11158,
  in which a series of strong flares
  emanated from its central PIL
  \citep{schr11}.
  \citet{tor14} and \citet{fang15}
  suggested that
  this AR is created
  by the emergence
  of a single flux tube
  that rises at two locations.
\item[Inter-AR] Strong flares
  produced on the PIL
  between two separated, apparently independent ARs.
  They show no clear $\delta$-configuration
  nor a clear sheared PIL
  at the flaring sites.
  The X1.2-class flare
  that occurred
  between NOAA ARs 11944 and 11943
  is a representative event
  of this category.
  It produced a very fast CME
  ($\sim 2,400\ {\rm km\ s}^{-1}$)
  that could potentially cause
  a severe geomagnetic disturbance
  \citep{moe15,wan15}.
\end{description}

The creation of non-potential structures
such as $\delta$-spots and sheared PILs
is a result of large-scale flux emergence
from the solar interior
and eventual sunspot motions.
In this paper,
in order to investigate the formation of 
the four above-mentioned types of ARs
that can potentially produce flares, CMEs,
and perhaps Earth-affecting disturbances,
we conduct a series of flux emergence simulations.
While flux emergence occurs
as a result of the dynamo mechanism
acting inside the Sun \citep{par55},
here we focus more on
the complexity and interaction
of magnetic flux systems
rising in the interior
and the resultant formation of ARs.

Flux emergence simulations
from the convection zone
have widely been used
to model solar ARs
in the last two decades
(e.g., \citealp{fan01,arc04,che10,tor11,tor12,rem14}:
for review see \citealp{che14}).
In the present work,
we test four different flux emergence simulations,
including those suggested previously
to model $\delta$-spots
\citep{fan99,tor14,fang15,tak15},
using similar numerical conditions,
and explore, in particular,
the formation of $\delta$-spots with sheared PILs
in the surface layer
as well as the buildup
of free magnetic energy
in the atmosphere.

Thanks to recent progress
in accurate magnetic measurements
and high-performance computations,
several flare and CME prediction methods
have been suggested and developed
\citep[e.g.,][]{lek03,sch07,wel09,fis12}.
\citet{bob15} extracted
various photospheric parameters
from vector magnetograms
(SHARP parameters)
and obtained good predictive performance
for $\geq$ M1.0 flares
using a machine learning algorithm.
In this paper,
we utilize
a series of numerical simulations,
which reproduce flaring ARs
with non-potential structures
($\delta$-spots and sheared PILs),
to examine
why these photospheric parameters
predict flare events
that occur in the corona
with higher accuracy.

The rest of the paper
is organized
as follows.
In Section \ref{sec:setup},
we describe the numerical setup
and assumed conditions
for the four simulation cases.
Then, in Section \ref{sec:general},
the general evolution
of the four cases
is shown.
Sections \ref{sec:shear} and \ref{sec:corona}
are dedicated to presenting
the detailed development
of $\delta$-spots and sheared PILs
in the photosphere
and the coronal energy buildup,
respectively,
while Section \ref{sec:sharp}
explores the prediction
of flares and CMEs
using photospheric parameters.
We summarize and discuss the results
in Sections \ref{sec:summary}
and \ref{sec:discussion},
respectively.

\section{Numerical Setup}\label{sec:setup}

\subsection{Assumptions and Basic Equations}\label{sec:setup:basic}

In this paper,
we investigated the emergence
of buoyant flux tubes
initially set in the convection zone.
We considered a rectangular computational domain
with three-dimensional (3D) Cartesian coordinates
$(x, y, z)$,
where the $z$-coordinate increases upward.
We solved the standard set
of resistive magnetohydrodynamic (MHD) equations:
\begin{eqnarray}
  \frac{\partial\rho}{\partial t}
  +\mbox{\boldmath $\nabla$}
  \cdot (\rho\mbox{\boldmath $V$})
  =0,
\end{eqnarray}
\begin{eqnarray}
  \frac{\partial}{\partial t}
  (\rho\mbox{\boldmath $V$})
  -\mbox{\boldmath $\nabla$}\cdot
  \left[
    \rho\mbox{\boldmath $V$}\mbox{\boldmath $V$}
    +\left(
    p+\frac{\mbox{\boldmath $B$}^{2}}{8\pi}
    \right){\bf I}
    -\frac{\mbox{\boldmath $B$}\mbox{\boldmath $B$}}{4\pi}
  \right]
  +\rho\mbox{\boldmath $g$}
  =0,
  \nonumber\\
\end{eqnarray}
\begin{eqnarray}
  \frac{\partial \mbox{\boldmath $B$}}{\partial t}
  =c\mbox{\boldmath $\nabla$}\times\mbox{\boldmath $E$},
\end{eqnarray}
\begin{eqnarray}
  && \frac{\partial}{\partial t}
  \left(
    \rho U
    +\frac{1}{2}\rho \mbox{\boldmath $V$}^{2}
    +\frac{\mbox{\boldmath $B$}^{2}}{8\pi}
  \right)
  \nonumber\\
  && +\mbox{\boldmath $\nabla$}\cdot
  \left[
    \left(
    \rho U
    +p
    +\frac{1}{2}\rho \mbox{\boldmath $V$}^{2}
    \right)
    \mbox{\boldmath $V$}
    +\frac{c}{4\pi}
    \mbox{\boldmath $E$}\times\mbox{\boldmath $B$}
  \right] 
  \nonumber\\
  && -\rho\mbox{\boldmath $g$}\cdot\mbox{\boldmath $V$}
  =0,
\end{eqnarray}
and
\begin{eqnarray}
  U=\frac{1}{\gamma-1}
  \frac{p}{\rho},
\end{eqnarray}
\begin{eqnarray}
  \mbox{\boldmath $E$}
  =-\frac{1}{c}
  \mbox{\boldmath $V$}\times\mbox{\boldmath $B$}
  +\eta\mbox{\boldmath $J$},
\end{eqnarray}
\begin{eqnarray}
  \mbox{\boldmath $J$}
  =\frac{c}{4\pi}
  \mbox{\boldmath $\nabla$}\times\mbox{\boldmath $B$},
  \label{eq:current}
\end{eqnarray}
\begin{eqnarray}
  p=\frac{k_{\rm B}}{m}\rho T,
\end{eqnarray}
where $\rho$ denotes the gas density,
$\mbox{\boldmath $V$}$ velocity vector,
$p$ pressure,
$\mbox{\boldmath $B$}$ magnetic field,
$c$ the speed of light,
$\mbox{\boldmath $E$}$ electric field,
and $T$ temperature,
while $U$ is the internal energy per unit mass,
{\bf I} the unit tensor,
$k_{\rm B}$ the Boltzmann constant,
$m (={\rm const.})$ the mean molecular mass,
and $\mbox{\boldmath $g$}=(0,0,-g_{0})=(0,0,-1/\gamma)$
the uniform gravitational acceleration.
We assumed the medium to be an inviscid perfect gas
with a specific heat ratio $\gamma=5/3$.

To make the above equations dimensionless,
we introduced the following normalizing units:
the pressure scale height $H_{0}=170\ {\rm km}$ for length,
the sound speed $C_{\rm s0}=6.8\ {\rm km\ s}^{-1}$ for velocity,
$\tau_{0}\equiv H_{0}/C_{\rm s0}=25\ {\rm s}$ for time,
and $\rho_{0}=1.4\times 10^{-7}\ {\rm  g\ cm}^{-3}$ for density,
all of which are typical values in the photosphere.
The gas pressure, temperature, and magnetic field strength were
normalized by the combinations of the units above,
i.e.,
$p_{0}=\rho C_{\rm s0}^{2}=6.3\times 10^{4}\ {\rm dyn\ cm}^{-2}$,
$T_{0}=mC_{\rm s0}^{2}/(\gamma k_{\rm B})=5,600\ {\rm K}$,
and $B_{0}=(\rho_{0}C_{\rm s0}^{2})^{1/2}=250\ {\rm G}$,
respectively.

We assumed an anomalous resistivity model
with the form
\begin{eqnarray}
  \eta=\left\{
  \begin{array}{ll}
    0 & (J<J_{\rm C}\ {\rm or}\ \rho>\rho_{\rm C})\\
    \eta_{0}(J/J_{\rm C}-1) & (J\geq J_{\rm C}\ {\rm and}\ \rho<\rho_{\rm C})
  \end{array}
  \right.,
\end{eqnarray}
where $\eta_{0}=0.1$,
$J_{\rm C}=0.1$,
and $\rho_{\rm C}=0.1$.
The above treatment is intended
to trigger magnetic reconnection
in a low-density current sheet.

\subsection{Numerical Conditions and the Reference Case}\label{sec:setup:initial}

\begin{figure*}
  \begin{center}
    \includegraphics[width=80mm]{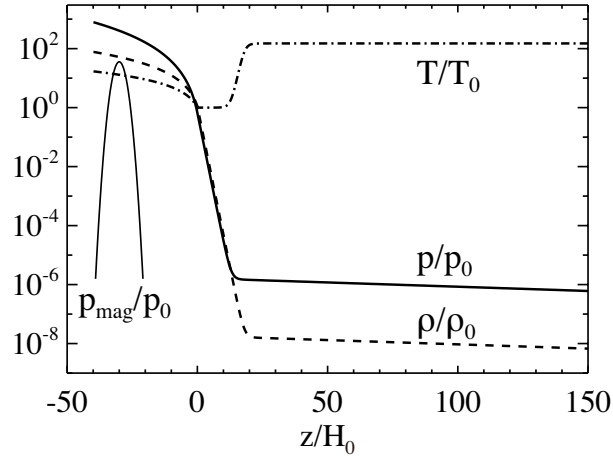}
  \end{center}
  \caption{One-dimensional ($z$-)distributions
    of the initial background density (thick solid),
    pressure (dashed),
    and temperature (dash-dotted).
    The magnetic pressure
    $p_{\rm mag}=B^{2}/(8\pi)$
    along the vertical axis $x=y=0$
    of the Reference case
    is overplotted (thin solid).
    \label{fig:initial}}
\end{figure*}

The initial background atmosphere
consisted of three regions:
an adiabatically stratified convection zone,
a cool isothermal photosphere/chromosphere,
and a hot isothermal corona
(see Figure \ref{fig:initial}).
We assumed $z/H_{0}=0$ to be
the base height of the photosphere,
and the initial temperature distribution
of the convection zone
($z/H_{0}\leq 0$)
was assumed to be
\begin{eqnarray}
  T(z)=T_{\rm ph}
  -z\left|
    \frac{dT}{dz}
  \right|_{\rm ad},
\end{eqnarray}
where $T_{\rm ph}=T_{0}$ is the photospheric temperature
and
\begin{eqnarray}
  \left|
    \frac{dT}{dz}
  \right|_{\rm ad}
  =\frac{\gamma-1}{\gamma}
  \frac{mg_{0}}{k_{\rm B}}
\end{eqnarray}
is the adiabatic temperature gradient;
i.e., the initial temperature profile
of the convection zone
is adiabatic.
The temperature distribution
of the atmosphere
($z/H_{0}\geq 0$)
was expressed as
\begin{eqnarray}
  T(z)=T_{\rm ph}
  +\frac{1}{2}
  (T_{\rm cor}-T_{\rm ph})
  \left[
    \tanh{
      \left(
        \frac{z-z_{\rm cor}}{w_{\rm tr}}
      \right)+1
    }
  \right],
  \nonumber\\
\end{eqnarray}
where $T_{\rm cor}=150T_{0}$ is the coronal temperature,
$z_{\rm cor}=18H_{0}$ is the base of the corona,
and $w_{\rm tr}=2H_{0}$ is the temperature scale height
of the transition region.
With the temperature profile above,
the initial pressure and density profiles
(Figure \ref{fig:initial})
were defined
by the equation of static pressure balance:
\begin{eqnarray}
  \frac{dp(z)}{dz}+\rho(z)g_{0}=0.
\end{eqnarray}

A magnetic flux tube was embedded
in the convection zone
and its longitudinal and azimuthal components
of the flux tube are given by
\begin{eqnarray}
  B_{x}(r)=B_{\rm tube}
  \exp{
    \left(
      -\frac{r^{2}}{R_{\rm tube}^{2}}
    \right)
  }
\end{eqnarray}
and
\begin{eqnarray}
  B_{\phi}=qrB_{x}(r),
\end{eqnarray}
where $r=[(y-y_{\rm tube})^{2}+(z-z_{\rm tube}^{2})]^{1/2}$
is the radial distance
from the tube axis,
$(y_{\rm tube}, z_{\rm tube})$
the location of the tube axis,
$R_{\rm tube}$ the radius,
$B_{\rm tube}$ the magnetic field strength at the axis,
and $q$ the twist intensity.
As the Reference (typical) case,
we considered $(y_{\rm tube}/H_{0}, z_{\rm tube}/H_{0})=(0,-30)$,
$R_{\rm tube}/H_{0}=3$,
$B_{\rm tube}/B_{0}=30$,
and $qH_{0}=-0.2$.
The total axial magnetic flux
amounts to
$\Phi_{\rm tube}/(B_{0}H_{0}^{2})=845$.
These parameters indicate that
the initial flux tube
is located at a depth of 5.1 Mm
and has a radius of 510 km;
a central field strength of 7.5 kG
(or the plasma $\beta\equiv8\pi p/B^{2}\sim 10$),
which yields an axial flux
of $6\times 10^{19}\ {\rm Mx}$;
and a left-handed twist.
The magnetic pressure,
$p_{\rm mag}=B^{2}/(8\pi)$,
along the vertical axis
is shown in Figure \ref{fig:initial}.
The gas pressure inside the tube
was defined as
$p_{\rm i}=p(z)+\delta p_{\rm exc}$,
with the pressure excess being
\begin{eqnarray}
  \delta p_{\rm exc}
  =\frac{B_{x}^{2}(r)}{8\pi}
  \left[
    q^{2}
    \left(
      \frac{R^{2}_{\rm tube}}{2}-r^{2}
    \right)-1
  \right] (<0).
  \label{eq:deltap}
\end{eqnarray}
To trigger the buoyant emergence,
we reduced the density
inside the flux tube,
$\rho_{\rm i}=\rho(z)+\delta\rho_{\rm exc}$,
where
\begin{eqnarray}
  \delta\rho_{\rm exc}
  =\rho(z)
  \frac{\delta p_{\rm exc}}{p(z)}
  \left[
    (1+\epsilon)
    \exp{
      \left(
        -\frac{(x-x_{\rm tube})^{2}}{\lambda^{2}}
      \right)
    }-\epsilon
  \right]
  \nonumber\\
  \label{eq:ptb}
\end{eqnarray}
and $x_{\rm tube}$ and $\lambda$
are the center and the length
of the buoyant section, respectively,
and $\epsilon$ is
a factor that suppresses the emergence
of both ends of the tube.
The typical values are
$x_{\rm tube}/H_{0}=0$,
$\lambda/H_{0}=8$,
and $\epsilon=0.2$.
Depending on the simulation case,
some parameters were modified
as described
in the next subsection.

The simulation domain was
$(-150, -150, -40)\leq (x/H_{0}, y/H_{0}, z/H_{0})\leq (150, 150, 400)$,
resolved by a $512\times 512\times 512$ grid.
The grid spacings
for the $x$-, $y$-, and $z$-directions were
$\Delta x/H_{0}=\Delta y/H_{0}=0.25$
for $(-20, -20)\leq (x/H_{0}, y/H_{0})\leq (20,20)$
and $\Delta z/H_{0}=0.2$
for $-40\leq z/H_{0}\leq 15$.
Outside this range,
the spacings were smoothly increased up to
$\Delta x/H_{0}=\Delta y/H_{0}=0.8$
and $\Delta z/H_{0}=1.8$.
We assumed a periodic boundary condition
for the $x$-direction
and symmetric boundaries
for both the $y$- and $z$-directions.

The simulation code we used
is the same as that used by \citet{tak15},
which is based on the numerical scheme
of \citet{voe05}:
4th-order central differences
for calculating the spatial derivatives
and the four-step Runge-Kutta scheme
for calculating the temporal derivatives.
Artificial diffusivity,
proposed by \citet{rem09},
was introduced to stabilize the calculation,
while the $\mbox{\boldmath $\nabla$}\cdot\mbox{\boldmath $B$}$ error
was reduced by the iterative hyperbolic divergence cleaning technique
based on the method described in \citet{ded02}.

\begin{figure*}
  \begin{center}
    \includegraphics[width=90mm]{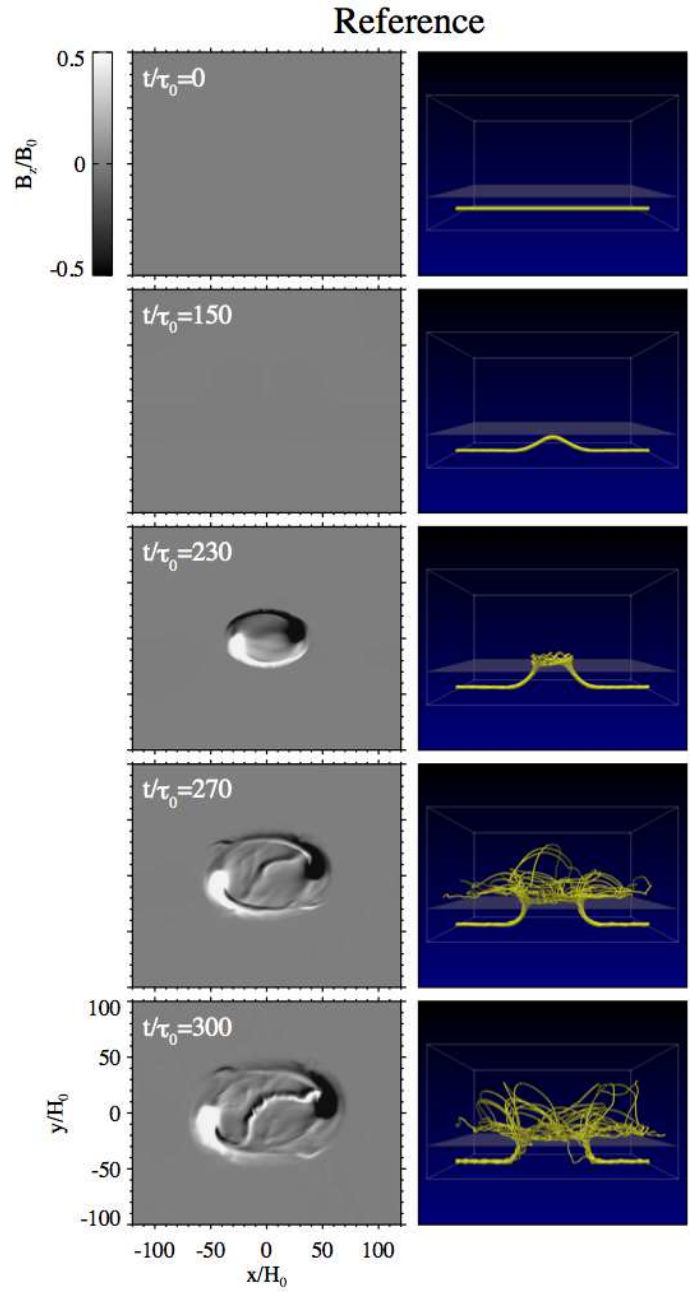}
  \end{center}
  \caption{Evolution of surface vertical magnetic fields
    and magnetic field lines
    for the Reference case.
    See the accompanying video
    for the temporal evolution.
    \label{fig:general_z}}
\end{figure*}

Figure \ref{fig:general_z} shows the evolution
of the magnetogram at $z/H_{0}=0$
and magnetic field lines
for the Reference case
(movie is attached to provide detailed evolution).
The whole evolution is consistent with
the previous 3D simulations
by, e.g., \citet{fan01}, \citet{arc04}, and \citet{tor12}:
the horizontal flux tube makes
an $\Omega$-shaped arcade,
which rises through the convection zone
and eventually penetrate the photosphere,
creating a magnetic dome
in the corona
with bipolar spots
in the photosphere.

\subsection{Four Simulation Cases}\label{sec:cases}

In order to model
the four types
of flare-productive ARs
introduced in Section \ref{sec:intro},
we tested four simulation cases
with initial conditions
different from those of the Reference case,
which are summarized
in the bottom row
of Figure \ref{fig:model}.

For the Spot-Spot case,
the initial twist strength
was intensified to
$qH_{0}=-0.8$,
which is larger
than the critical value
for the kink instability
($|q|H_{0}=0.33$: \citealt{lin96}).
Due to the stronger initial twist,
the density deficit is larger
for this case
(Equations (\ref{eq:deltap}) and (\ref{eq:ptb}))
and thus the flux tube starts
with a faster rising speed
\citep[see, e.g.,][]{mur06}.
At the same time,
the kinking itself accelerates
the flux tube:
When the tube kinks,
its axis is stretched,
which enhances the buoyancy
and makes the rise speed faster
\citep{fan99}.

The second case,
Spot-Satellite,
was modeled by introducing
a parasitic flux tube
set in a direction
perpendicular to the main flux tube.
Perhaps this type can also be produced
from a single flux tube
that bifurcates.
However, in this paper,
for simplicity,
we tested the two-tube scenario
(main and parasitic tubes).
The parasitic tube has parameters
of $R_{\rm tube}/H_{0}=2$,
$B_{\rm tube}/B_{0}=15$,
and $qH_{0}=-0.2$
(directed to $+y$ with a left-handed twist).
The tube center is located
at $(x/H_{0}, y/H_{0}, z/H_{0})=(15, 0, -14)$
and is kept in mechanical balance.
In this model,
a periodic boundary condition
was applied in the $y$-direction.

The Quadrupole flux tube
has two buoyant sections
along the axis
and thus starts emergence
at two locations.
We changed the density perturbation
of Equation (\ref{eq:ptb}) to
\begin{eqnarray}
  \delta\rho_{\rm exc}
  =\rho(z)
  \frac{\delta p_{\rm exc}}{p(z)}
  \left[
  (1+\epsilon)
  \exp{
    \left(
      -\frac{(x-x_{\rm tube1})^{2}}{\lambda^{2}}
    \right)
  }
  \right. 
  \nonumber\\
  \left. 
  +(1+\epsilon)
  \exp{
    \left(
      -\frac{(x-x_{\rm tube2})^{2}}{\lambda^{2}}
    \right)
  }
  -\epsilon
  \right],
\end{eqnarray}
where $x_{\rm tube1}/H_{0}=-3\lambda/H_{0}=-24$
and $x_{\rm tube2}/H_{0}=3\lambda/H_{0}=24$.

Finally,
for the Inter-AR case,
we set two flux tubes in parallel
in the convection zone.
The two tubes have parameters of
$(x_{\rm tube1}/H_{0}, y_{\rm tube1}/H_{0})=(3\lambda/H_{0},-3\lambda/H_{0})=(24,-24)$
and $(x_{\rm tube2}/H_{0}, y_{\rm tube2}/H_{0})=(-3\lambda/H_{0},3\lambda/H_{0})=(-24,24)$.

In this work,
for the purposes
of comparing the simulations to observations,
we refer to the emerged region
as $\delta$-spots
if it is complex (qualitatively),
compact (separation of the both polarities
less than, say, $20H_{0}$),
and highly sheared
(shear angle of the PIL $\sim 90^{\circ}$).
Note that we take these values for
just a threshold in the simulations
and they are not actually measured from observations.
Also, although previous observations found that
the $\delta$-spots
often show rotational motions
and violate Hale's polarity rule
\citep[e.g.][]{kur87,lop00,lop03},
these properties are not used
as the definition here.

\section{General Evolution}\label{sec:general}

\begin{figure*}
  \begin{center}
    \includegraphics[width=90mm]{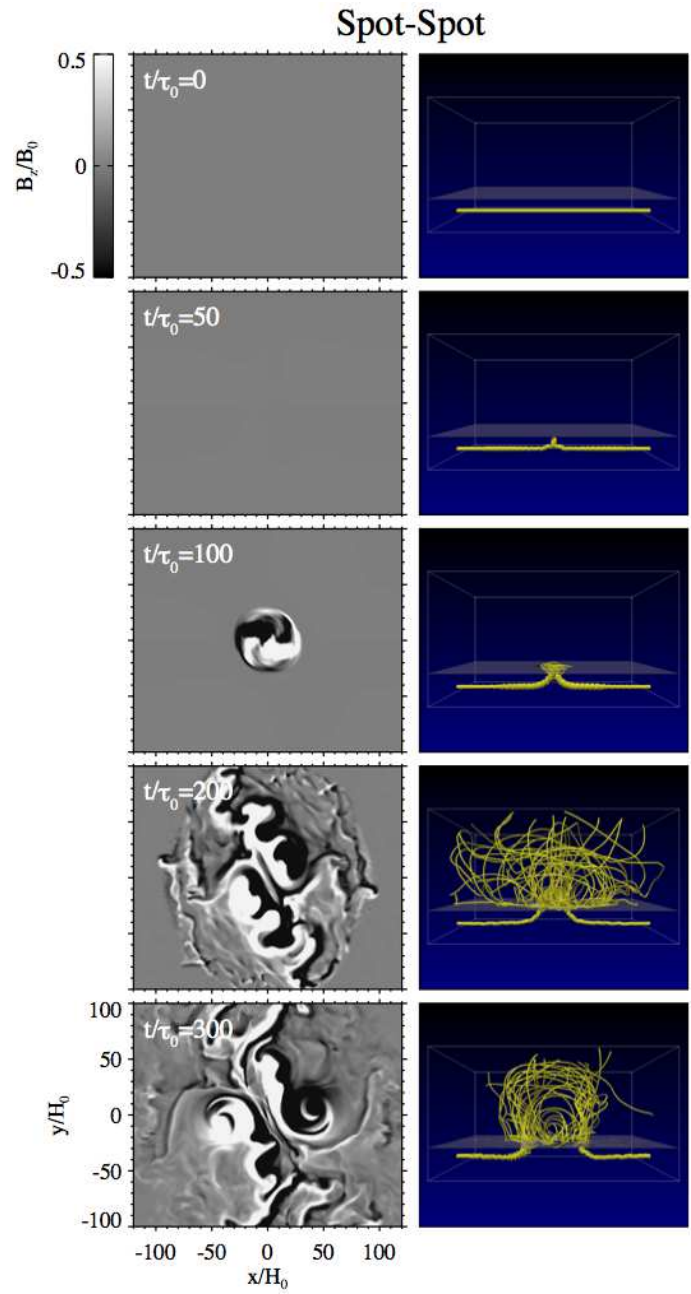}
  \end{center}
  \caption{Same as Figure \ref{fig:general_z}
    but for the Spot-Spot case.
    \label{fig:general_a}}
\end{figure*}

\begin{figure*}
  \begin{center}
    \includegraphics[width=90mm]{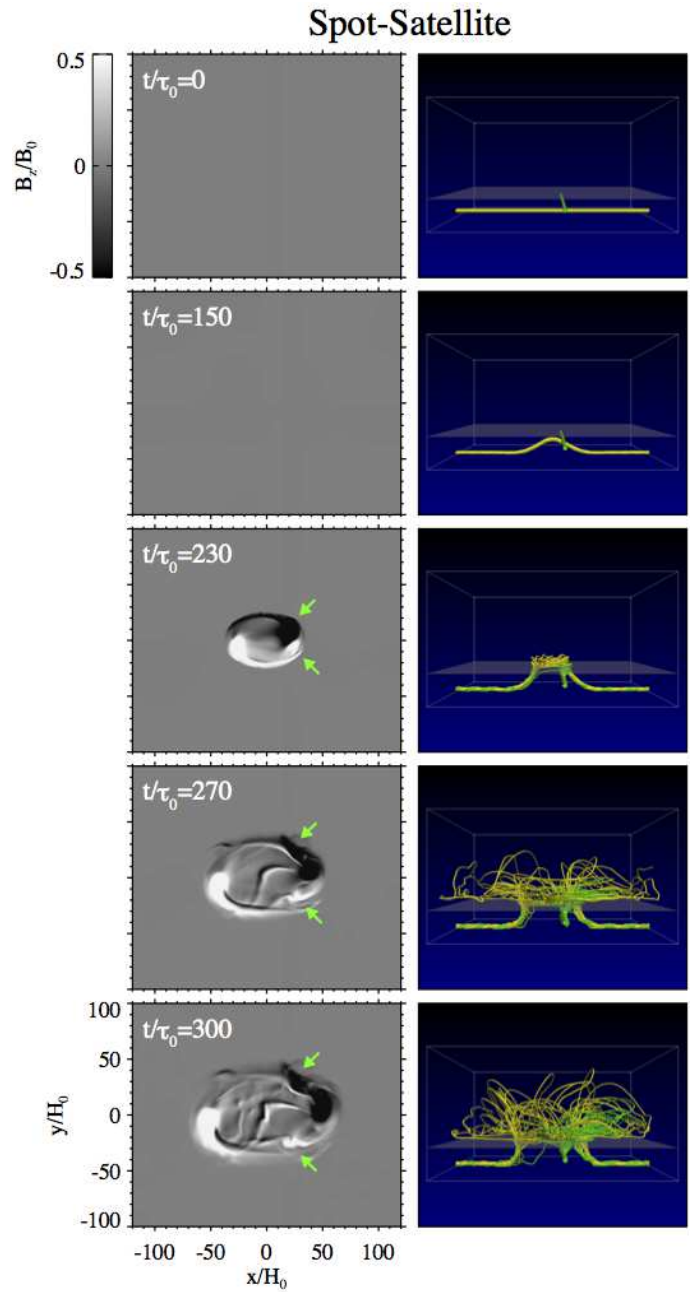}
  \end{center}
  \caption{Same as Figure \ref{fig:general_z}
    but for the Spot-Satellite case.
    The green arrows in the magnetograms
    indicate the satellite spots,
    which originate
    from the parasitic flux tube,
    while the green field lines
    in the right column
    are for the parasitic tube.
    \label{fig:general_b}}
\end{figure*}

\begin{figure*}
  \begin{center}
    \includegraphics[width=90mm]{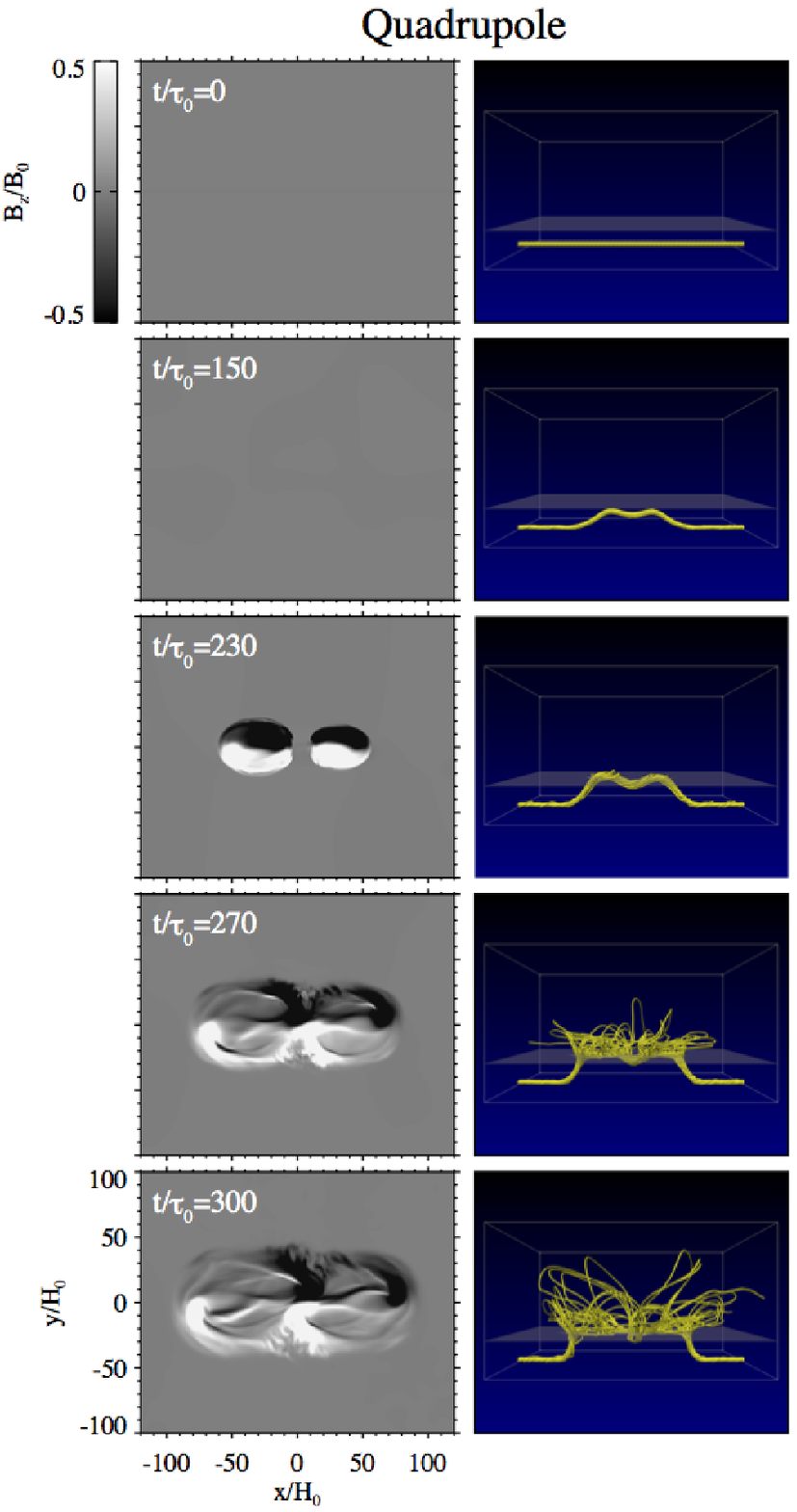}
  \end{center}
  \caption{Same as Figure \ref{fig:general_z}
    but for the Quadrupole case.
  \label{fig:general_c}}
\end{figure*}

\begin{figure*}
  \begin{center}
    \includegraphics[width=90mm]{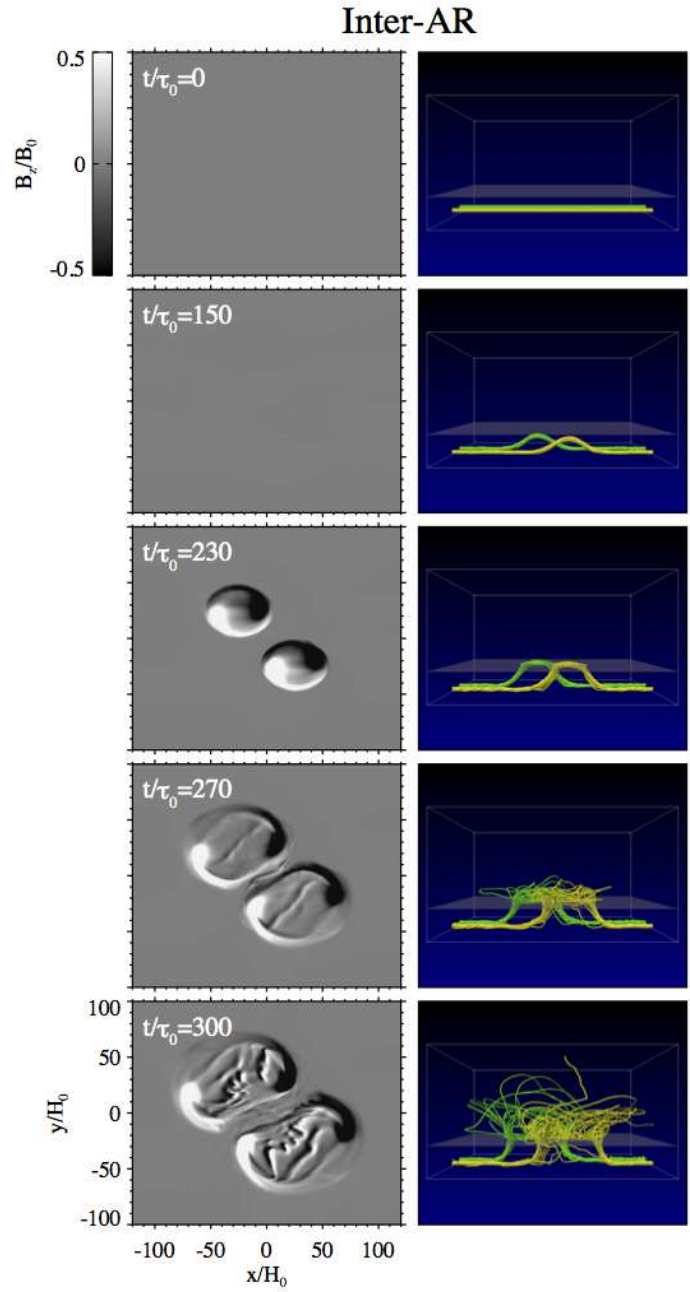}
  \end{center}
  \caption{Same as Figure \ref{fig:general_z}
    but for the Inter-AR case.
    The green field lines
    are for the secondary tube.
  \label{fig:general_d}}
\end{figure*}

\begin{figure*}
  \begin{center}
    \includegraphics[width=80mm]{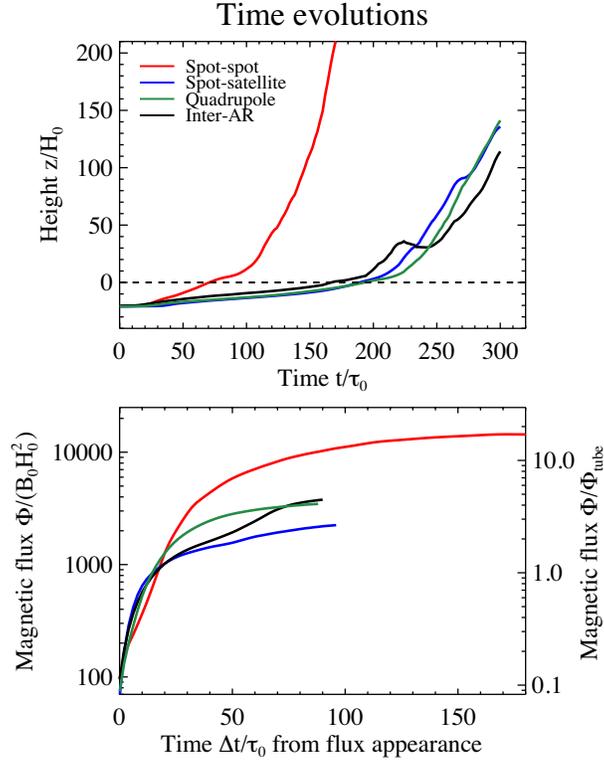}
  \end{center}
  \caption{(Top) Time evolution
    of the highest part of the flux tubes
    for the four simulation cases.
    Solar surface ($z/H_{0}=0$) is indicated
    by a dashed line.
    For the Spot-Satellite case,
    only the contribution of the main tube
    is shown.
    (Bottom) Evolution of the total unsigned magnetic flux,
    $\Phi=\int |B_{z}|dx\,dy$,
    measured at the solar surface.
    Time $\Delta t/\tau_{0}$ is measured
    since the flux appears
    at the surface.
    The left vertical axis indicates
    the non-dimensional value
    of the magnetic flux,
    i.e., in the unit of $B_{0}H_{0}^{2}$,
    while the right vertical axis presents
    the value normalized
    by the total axial magnetic flux
    of the initial flux tube,
    $\Phi_{\rm tube}/(B_{0}H_{0}^{2})=845$.
    \label{fig:evolution}}
\end{figure*}

Figures \ref{fig:general_a}, \ref{fig:general_b},
\ref{fig:general_c}, and \ref{fig:general_d}
are the photospheric magnetograms
and field lines
for the Spot-Spot, Spot-Satellite,
Quadrupole, and Inter-AR cases,
respectively,
while Figure \ref{fig:evolution} compares
the apex heights, $z/H_{0}$,
as a function of $t/\tau_{0}$
and the surface total unsigned magnetic flux,
$\Phi=\int |B_{z}|\, dx\, dy$,
since the flux appears at the surface.
From these diagrams,
one can observe that
the most drastic evolution
is for the Spot-Spot case,
i.e., the emergence of a highly-twisted kink-unstable flux tube.
As explained in detail
in earlier works
by \citet{fan99} and \citet{tak15},
this flux tube quickly develops
a knotted structure
in the convection zone
rather than making a simple $\Omega$-shaped arch
(see field line rendering at $t/\tau_{0}=50$),
reducing its twist about the axis
by making the axis writhed.
It reaches the surface around $t/\tau_{0}=75$
with the orientation of the axis
that connects the two main surface polarities
highly deviated from the direction
of the original flux tube
(see magnetogram at $t/\tau_{0}=100$);
i.e., this AR violates
Hale's polarity rule.
Eventually,
at $t/\tau_{0}=300$,
the magnetogram shows a pair
of circular spots
of opposite polarities
around $(x/H_{0},y/H_{0})=(\pm 45, \pm 5)$
with extended tails.
An elongated PIL
is built
in the middle of the domain,
sandwiched by the two main sunspots.
The total magnetic flux at the surface
is more than 10 times
the original axial flux.
This is mostly because
the original flux tube has a strong twist
and thus a large amount of azimuthal flux
in addition to the axial component,
but this is also because
some field lines wander up and down
the surface layer,
which increases
the total unsigned flux.
Since this AR is composed of bipolar spots
with scattered patches
and closely-neighboring opposite polarities,
it can be classified as a $\beta\gamma\delta$ spot.

The remaining three cases show
relatively gentle evolutions.
For the Spot-Satellite case
(Figure \ref{fig:general_b}),
the rising $\Omega$-shaped main tube
comes into contact
with the resting parasitic tube
at $t/\tau_{0}=150$
and starts pushing it up.
From $t/\tau_{0}=200$,
the surface magnetogram shows
a separation of the main bipolar spots
in the $x$-direction,
with minor satellite spots
separating along the $y$-axis
at $x/H_{0}\sim 25$
(see green arrows in the magnetogram).
A compact PIL is formed
between the negative main spot
and the positive satellite polarity
only for a short period
when the positive polarity
transits alongside the negative spot
and forms a $\delta$-spot structure.
In the corona,
field lines of the parasitic tube
(green lines in field line rendering)
are pushed aside
along the positive $x$-direction
by the main flux tube (yellow lines).
Owing to magnetic reconnection,
some green field lines
trace the original tube
in the convection zone
deeper down to both footpoints.
The final surface flux is
slightly larger than twice
the original axial flux,
probably because of the contribution
of the satellite spots
(Figure \ref{fig:evolution}).
We conjecture that
this AR can also be a $\beta\gamma\delta$ spot.

The two density-deficit sections
along the flux tube
of the Quadrupole case
create an M-shaped configuration
in the convection zone
(Figure \ref{fig:general_c}).
The rising speed in the interior
is approximately the same
as that of the Spot-Satellite case.
From $t/\tau_{0}=200$,
the flux tube creates
a pair of bipoles
at the surface,
and from $t/\tau_{0}=250$,
the two central polarities,
tightly connected by the dipped field lines
beneath the surface,
collide against each other,
forming
a closely-packed ($\delta$-like) sunspot
with a clearly defined PIL.
The field lines show
two expanded magnetic domes
in the corona.
The final surface flux is about four times
the original axial values
(Figure \ref{fig:evolution}),
which indicates that the flux tube
comes in and out of the surface twice.
The above process is in good agreement
with previous simulations
\citep{tor14,fang15}.
Probably this AR can be categorized as
$\beta\delta$ or $\beta\gamma\delta$.

The two flux tubes
of the Inter-AR case
follow a similar development process
(Figure \ref{fig:general_d}).
However,
since the two inner polarities
are not connected
by the subsurface field lines,
they have almost no contact with each other
and simply show a fly-by motion.
Consequently,
a strong field-gradient PIL is not created
in this case.
The final value
of the surface magnetic flux
is about four times
the initial axial flux
(Figure \ref{fig:evolution}),
which is again a behavior
similar to the Quadrupole case.
The above evolution
is consistent with Case 2
of \citet{tor14}.
Contrary to the previous cases,
the two ARs in this simulaiton
should be simply regarded as $\beta$-spots.

\section{Formation of $\delta$-spots and Sheared PILs}\label{sec:shear}

As discussed in Section \ref{sec:intro},
sheared magnetic structures
are thought to be important
for the production of strong flare events.
In particular, the sheared PIL
in a $\delta$-shaped sunspot
is one of the most preferable locations
for flare production.
In this section,
we show the detailed formation processes
of $\delta$-spots with sheared PILs
for the four simulation cases.

\subsection{Spot-Spot}

\begin{figure*}
  \begin{center}
    \includegraphics[width=150mm]{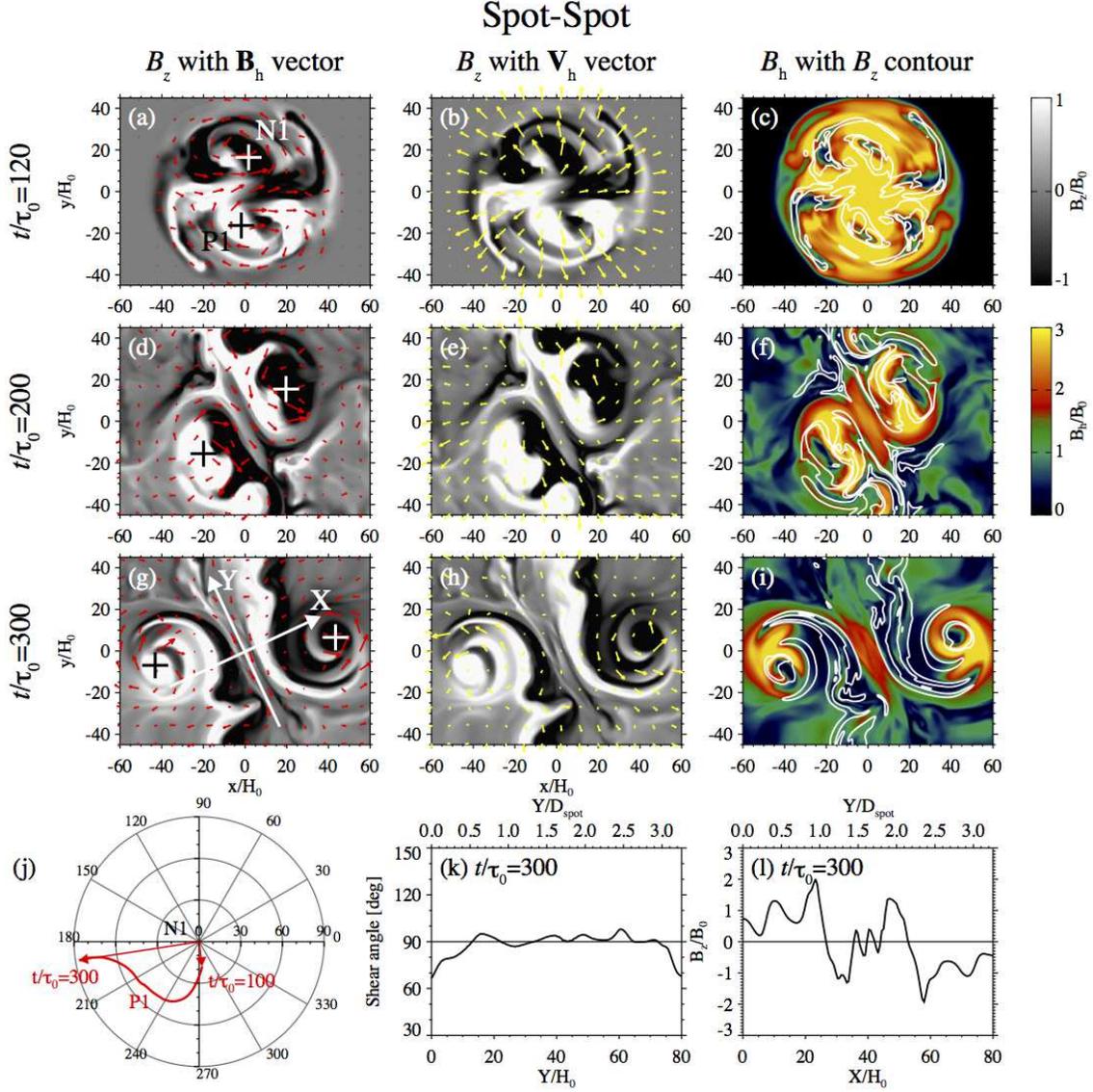}
  \end{center}
  \caption{Formation of $\delta$-spot and sheared PIL
    for the Spot-Spot case.
    The left column shows the magnetogram
    ($B_{z}$: black-white)
    with horizontal magnetic fields
    ($\mbox{\boldmath $B$}_{\rm h}$: red arrows)
    at three different times.
    Plus signs denote
    the centers of the two main sunspots
    of positive (P1) and negative (N1) polarities,
    which are defined as the local maximum and minimum
    of the vertical fields,
    respectively.
    The local coordinates $(X, Y)$ in panel (g)
    are defined such that the $Y$-axis is parallel
    to the developed PIL.
    The horizontal velocity
    ($\mbox{\boldmath $V$}_{\rm h}$: yellow arrows)
    is shown in the middle column,
    while in the right column,
    the horizontal field strength
    ($B_{\rm h}$: color) is presented.
    Panel (j) shows the relative motion
    of the sunspots N1 and P1.
    The center of the diagram corresponds to N1,
    the horizontal axis is parallel to the $x$-axis,
    and the arrow head indicates
    the relative position of P1.
    Panels (k) and (l) are
    the physical parameters
    along the $X$ and $Y$ axes
    in panel (g):
    the shear angle,
    $\arctan{(B_{Y}/B_{X})}$,
    along the $Y$-axis
    and the vertical field,
    $B_{z}/B_{0}$,
    along the $X$-axis.
    For comparison,
    a length scale
    normalized by the typical sunspot diameter,
    $D_{\rm spot}/H_{0}=24.6$,
    is also shown
    as the upper horizontal axis
    (see main text for details).
    \label{fig:pil_test061}}
\end{figure*}

Figure \ref{fig:pil_test061} summarizes
the detailed photospheric evolution
of the Spot-Spot case.
In this case,
as the twisted flux tube emerges,
a complex magnetic pattern is formed
in the surface layer
and eventually an elongated PIL,
highlighted by the $Y$-axis,
is created
at the center
between the sunspot pair,
P1 and N1.
One prominent feature here
is the counter-streaming shear flow
along the PIL
(see $\mbox{\boldmath $V$}_{\rm h}$ vector),
with its orientation following
the expansion of the magnetic arcades
in the atmosphere.
As a result of the shear flow,
the horizontal magnetic field
becomes highly inclined
to the PIL direction
(see $\mbox{\boldmath $B$}_{\rm h}$ vector)
and the shear angle
becomes almost $90^{\circ}$
(see panel (k)).
Here, the shear angle is measured
from the direction of the potential field
(the direction perpendicular to the PIL)
and thus $90^{\circ}$ is parallel to the PIL.
The length of the highly-sheared ($\sim 90^{\circ}$) part
along the PIL
is $L_{\rm PIL}/H_{0}\sim 60$.

For easy comparison of the PIL
with other simulation cases
and actual observations,
we introduce the diameter of a Reference sunspot.
In the Reference case
in Section \ref{sec:setup:initial},
the final photospheric magnetogram
at $t/\tau_{0}=300$
shows a simple bipolar pair
(Figure \ref{fig:general_z}).
We measure the area of this Reference spot
(region with $|B_{z}|/B_{0}\ge 0.5$)
and define $D_{\rm spot}$
as the diameter of the circle
with an area equivalent to
the area of this spot.
Then, we obtain $D_{\rm spot}=24.6$,
which is used as a normalizing factor
for length scale
in Figures \ref{fig:pil_test061}(k) and (l).
With this value,
one can find that
the length of the highly-sheared PIL
of the Spot-Spot case
is as large as $L_{\rm PIL}/D_{\rm spot}=2.5$.

One may also notice that
the developed PIL
has a strong horizontal field
(see panel (i)).
Moreover,
this PIL reveals
an alternating pattern
of positive and negative polarities
(see panels (h) and (l)).
These features are highly reminiscent
of the ``magnetic channel'' structure,
which is introduced by \citet{zir93}
as one of the key characteristics
of the flare-producing PIL
\citep{kub07,wan08}.
From the comparison with numerical simulations,
\citet{kus12} and \citet{bam13}
suggested the possibility that
small-scale flux emergence at the magnetic channel
in NOAA AR 10930
triggers the series of large flare eruptions.

\subsection{Spot-Satellite}

\begin{figure*}
  \begin{center}
    \includegraphics[width=150mm]{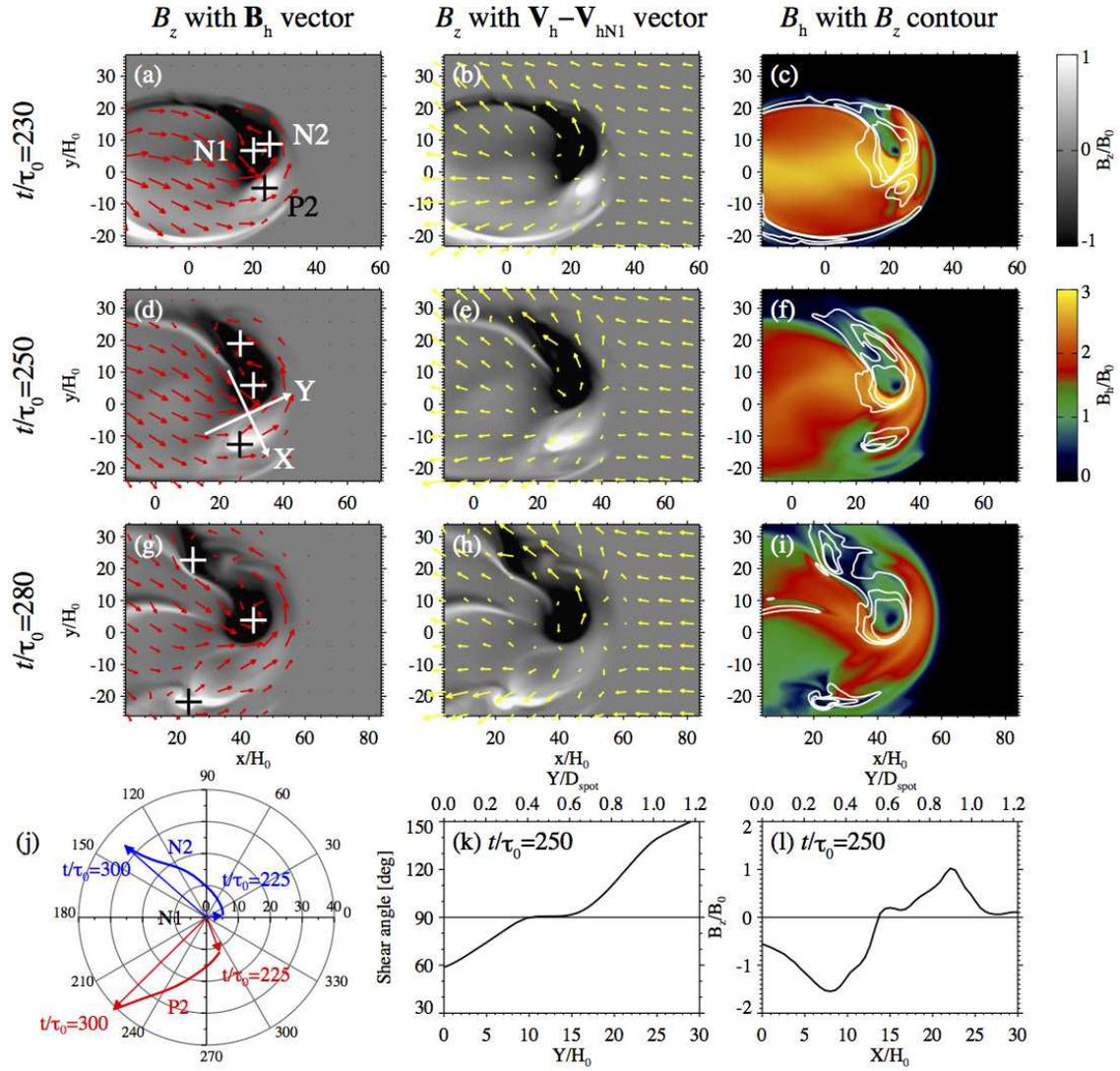}
  \end{center}
  \caption{Same as Figure \ref{fig:pil_test061}
    but for the Spot-Satellite case.
    Panels (a) to (i) are shown
    so that the negative main polarity N1
    is always located at the center of the diagram.
    In the middle column,
    the relative horizontal velocity,
    $\mbox{\boldmath $V$}_{\rm h}-\mbox{\boldmath $V$}_{\rm hN1}$,
    is plotted.
    In panel (j),
    the relative motion of N1 and N2
    is also shown.
    \label{fig:pil_test062}}
\end{figure*}

In the Spot-Satellite case,
the newly emerging field
in close proximity to the main sunspot
of the opposite polarity
creates a compact sheared PIL
within a $\delta$-spot.
Figure \ref{fig:pil_test062} shows that
the small bipole P2-N2 appears
immediately right of ($+x$-side of) the main spot N1.
The horizontal flow field and the relative motion
(middle column and panel (j)) indicate that,
as N1 proceeds to the right,
P2 drifts along the lower edge of ($-y$-side of) N1
and produces a sheared PIL.
Reflecting the scale
of the parasitic tube
and thus of the satellite spot,
the length of the highly-sheared part
of the PIL
at $t/\tau_{0}=250$
is only $L_{\rm PIL}/H_{0}\sim 5$
or about 20\% of the typical spot diameter,
$D_{\rm spot}$.
Furthermore,
in this case,
the horizontal field
is stronger at the PIL
(see panel (f)).

The newly emerging fields
at the edge of preexisting sunspots
have been reported
to drive
major flares occasionally.
For example,
\citet{lou14} found that
emerging satellite spots
ahead of the leading sunspot
of NOAA AR 11515
produce a filament
at the PIL,
which eventually erupts
at the onset of the M5.6-class flare
that develops into a CME:
compare especially their Figure 7
and Figure \ref{fig:pil_test062}
of this paper.
Similar behaviors have been reported
by, e.g., \citet{wan91}, \citet{ish98}, \citet{sch94}, and \citet{tak04},
while simulations have shown that
CME eruptions can be triggered
by newly emerged flux
at the edges of ARs
\citep[e.g.][]{che00}.

In this case,
as the satellite polarities (P2 and N2)
move away from the main spot (N1),
the $\delta$-configuration and sheared PIL
eventually disappear
(see $t/\tau_{0}=280$ in Figure \ref{fig:pil_test062}).
The $\delta$-spots are only seen
in the earliest phase
of the satellite emergence.

\subsection{Quadrupole}

Advected by horizontal flows
(middle column of Figure \ref{fig:pil_test063}),
the two inner sunspots
of the Quadrupole case,
N1 and P2,
collide with each other
at the center of the simulation domain.
The distance between the two spots
show a monotonic decrease
(panel (j))
and eventually
a strongly packed $\delta$-spot is created.
The highly-sheared PIL has a length
of $L_{\rm PIL}/H_{0}\sim 15$,
or $L_{\rm PIL}/D_{\rm spot}\sim 0.6$,
with the strongest $B_{z}$ gradient
(see panel (l)).
The horizontal field is best enhanced
at the central PIL
(panel (i)).

\citet{sun12} showed that
the Quadrupole AR NOAA 11158,
producing the first X-class event
in Cycle 24,
hosts a highly-sheared PIL
between the two colliding sunspots
\citep{schr11,liu12,tor13}.
They pointed out that
the PIL has a strong horizontal field,
which is in good agreement
with the PIL simulated here:
compare Figure 5 of \citet{sun12}
and Figure \ref{fig:pil_test063}
of this paper.

\subsection{Inter-AR}

The final case,
Inter-AR,
does not show
the clear formation
of a sheared PIL
nor a $\delta$-spot
(Figure \ref{fig:pil_test064}).
The two inner sunspots,
N1 and P2,
remain separated from each other
and simply show a fly-by motion
(see panel (j)).
In the central region in the photosphere,
the horizontal field exhibits
a slight indication
of a magnetic shear
(panel (g): $Y$-axis).
However,
the shear angle and the $B_{z}$ gradient
are not significant
(panels (k) and (l)).

As mentioned in Section \ref{sec:intro},
the X1.2-class flare
from NOAA ARs 11944 and 11943
produced a very fast CME.
\citet{moe15} pointed out
the importance of AR magnetic structures
in controlling the eruption of the CME.
Although this flare event
was not from the sheared PIL,
it produced a fast CME
that channeled through the open magnetic flux
created between the two closed field systems,
ARs 11944 and 11943.

\subsection{Factors That Contribute to the Development of Sheared PIL}

\begin{figure*}
  \begin{center}
    \includegraphics[width=150mm]{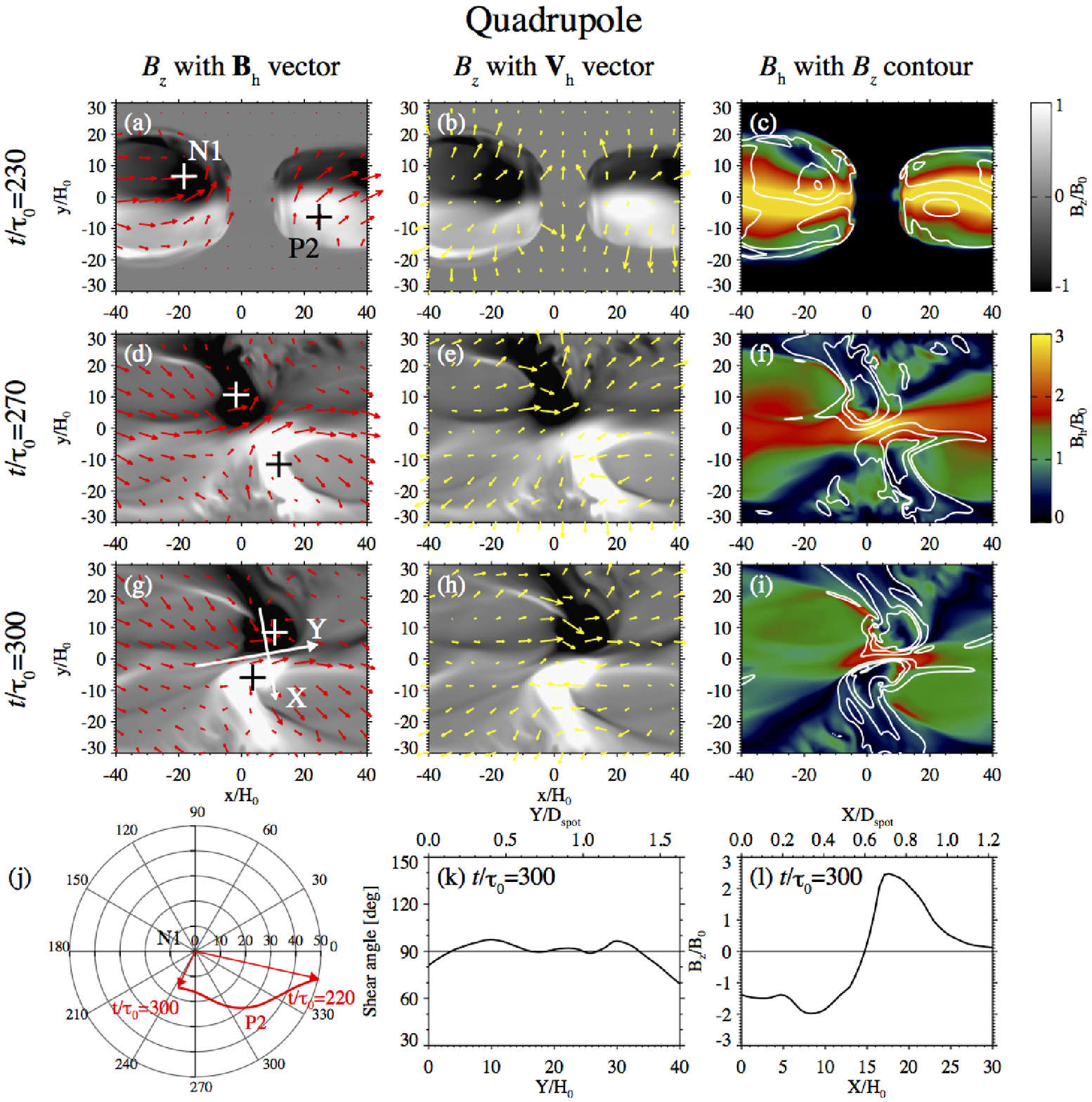}
  \end{center}
  \caption{Same as Figure \ref{fig:pil_test061}
    but for the Quadrupole case.
    \label{fig:pil_test063}}
\end{figure*}

\begin{figure*}
  \begin{center}
    \includegraphics[width=150mm]{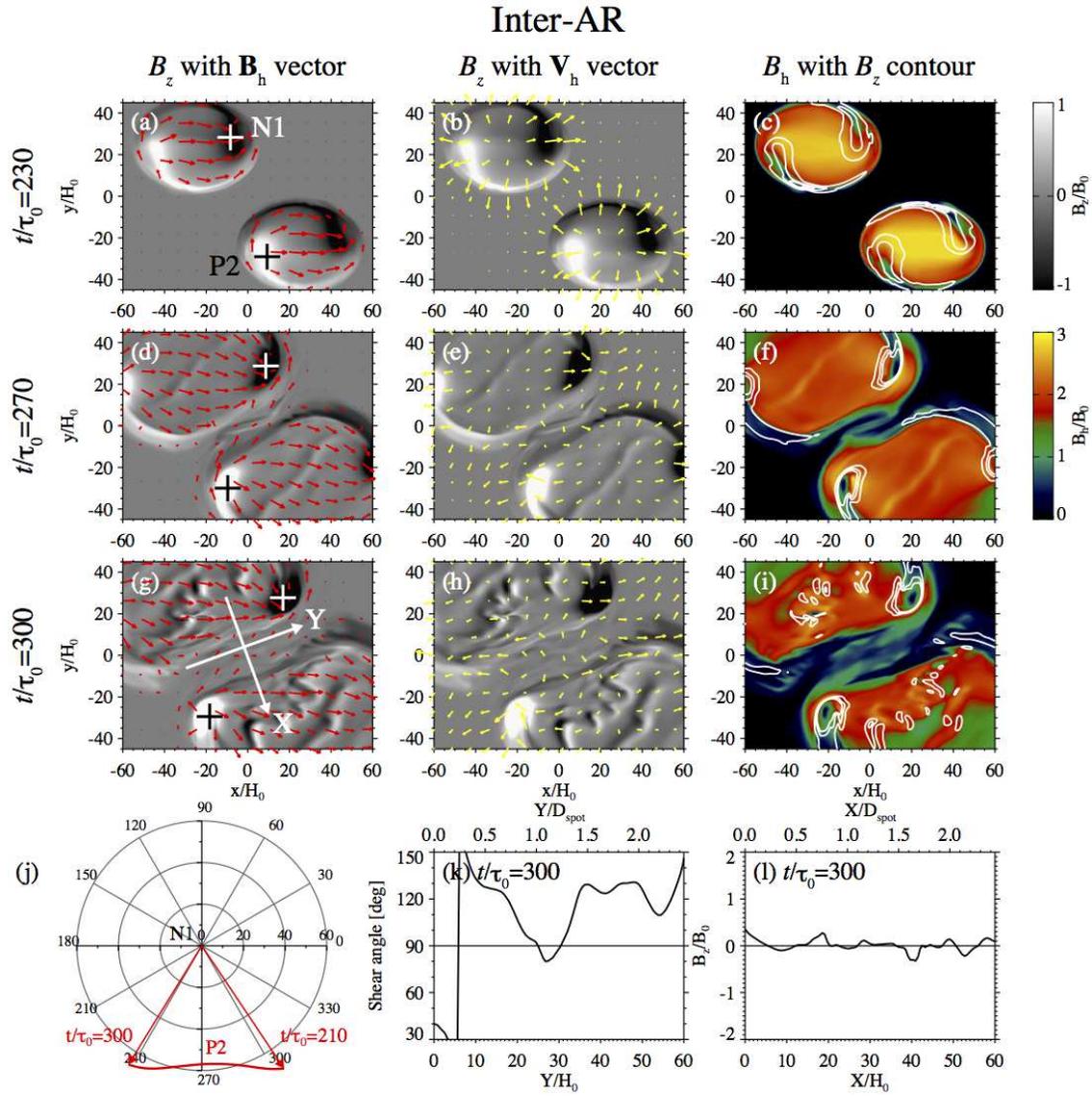}
  \end{center}
  \caption{Same as Figure \ref{fig:pil_test061}
    but for the Inter-AR case.
    \label{fig:pil_test064}}
\end{figure*}

\begin{figure*}
  \begin{center}
    \includegraphics[width=80mm]{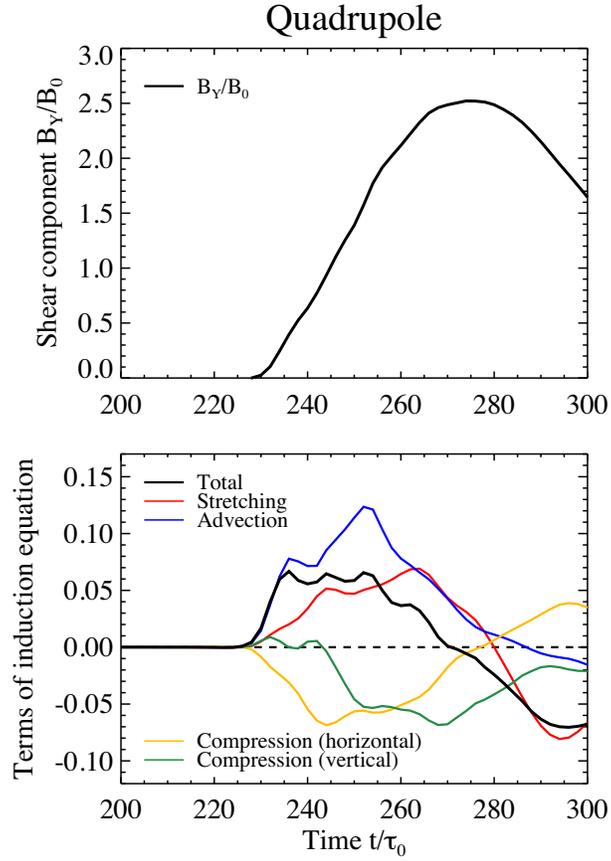}
  \end{center}
  \caption{Evolution of the magnetic shear
    at the PIL of the Quadrupole case.
    (Top) Time evolution
    of the shear component
    $B_{Y}/B_{0}$
    averaged over $15\le Y/H_{0}\le 25$,
    where the $Y$-axis is shown
    in Figure \ref{fig:pil_test063}(g).
    (Bottom) Evolution of each term
    of the induction equation:
    see Equation (\ref{eq:inductioneq}).
    The zero level is indicated
    by a horizontal dashed line.
    \label{fig:inductioneq}}
\end{figure*}

Among the four simulation cases
of flare-productive ARs,
we found that the Quadrupole case
produces the highly-sheared PIL
with the largest $B_{z}$ gradient
in a well developed $\delta$-spot.
In order to investigate
the evolution of the sheared PIL,
we take the Quadrupole case
as an example
and plot the terms of the induction equation
in Figure \ref{fig:inductioneq}.
Here
we show the shear component
of the photospheric horizontal field, $B_{Y}/B_{0}$,
i.e., the magnetic field
along the $Y$-axis
in Figure \ref{fig:pil_test063}(g),
and each term
of the induction equation,
\begin{eqnarray}
  \frac{\partial B_{Y}}{\partial t}
  = && \underbrace{(\mbox{\boldmath $B$}\cdot\mbox{\boldmath $\nabla$})V_{Y}}_{\rm Stretching}
  \underbrace{-(\mbox{\boldmath $V$}\cdot\mbox{\boldmath $\nabla$})B_{Y}}_{\rm Advection}
  \nonumber\\
  && \underbrace{-\left(\frac{\partial V_{X}}{\partial X}+\frac{\partial V_{Y}}{\partial Y}\right)B_{Y}}_{\rm Compression\, (horizontal)}
  \underbrace{-\frac{\partial V_{z}}{\partial z}B_{Y}}_{\rm Compression\, (vertical)}.
  \label{eq:inductioneq}
\end{eqnarray}
In Equation (\ref{eq:inductioneq}),
we neglect the magnetic diffusion
and divide the compression term,
$-(\mbox{\boldmath $\nabla$}\cdot\mbox{\boldmath $V$})B_{Y}$,
into the horizontal and vertical components.

It is seen from Figure \ref{fig:inductioneq} that
the shear field $B_{Y}/B_{0}$
appears
at $t/\tau_{0}=230$,
peaks around $t/\tau_{0}=275$,
and then gradually decays.
During this period,
the advection term is initially dominant
($t/\tau_{0}\lesssim 260$)
and, as the advection weakens,
the stretching term increases
and becomes comparable to the advection term
($260\lesssim t/\tau_{0}\lesssim 280$).
For most of the time,
the two compression terms remain negative
and do not contribute to the growth of the shear.
In the final phase
after $B_{Y}$ attains its peak
($t/\tau_{0}\gtrsim 280$),
the total value becomes negative
and thus $B_{Y}$ decreases.
However,
the horizontal compression turns positive
and becomes the only term
that works to sustain $B_{Y}$.

The above behavior
can be explained in the following manner.
In the Quadrupole case,
the sunspots of positive and negative polarities
(N1 and P2 in Figure \ref{fig:pil_test063})
approach to the region center
and produce
a $\delta$-like configuration.
In the early phase,
as the two spots come closer,
they transport the horizontal field
from both sides
(see the horizontal field vector
shown with red arrows
in Figure \ref{fig:pil_test063}).
This effect enhances the advection term
in the early phase (Figure \ref{fig:inductioneq}).
Then,
after the two spots merge,
they show a drifting motion
(N1 to the right and P2 to the left:
yellow arrows in Figure \ref{fig:pil_test063}),
which stretches
the horizontal field
along the PIL,
leading to the enhancement
of the stretching term
in the later phase
(Figure \ref{fig:inductioneq}).
The compression by the two approaching spots
also becomes stronger.
However,
this is only true for the horizontal component
(Figure \ref{fig:inductioneq}).
Since the emergence is a process
of a nonlinear instability,
the rising field drastically expands vertically
and $\partial V_{z}/\partial z$ is positive
\citep[e.g.,][]{shi89}.
The negative contribution
of the vertical compression term
in Figure \ref{fig:inductioneq}
reflects this process.

\section{Magnetic Structures and Energy Buildup in the Atmosphere}\label{sec:corona}

\subsection{Magnetic Structures}

Figure \ref{fig:fl} summarizes
the 3D magnetic field structures
for the four simulation cases.
For the Spot-Spot case,
the green magnetic field lines,
each connecting the main spot and the extended tail,
approach the center of the domain
from both sides
and form an electric current sheet
between them
(indicated by an isosurface
in the middle column:
see Appendix \ref{sec:current}
for the plotted values).
As a result,
the green field lines
reconnect with each other
and create the purple and yellow field lines.
The newly created purple flux system
is highly sheared
and aligned almost parallel
to the photospheric PIL.
However,
this purple flux is trapped
by the overlying yellow flux
that connects the two main sunspots.

In the Spot-Satellite case,
as the main flux tube (yellow)
pushes up the parasitic tube (green),
magnetic reconnection
occurs between the two flux tubes
and the purple field lines are formed.
One may find that,
in the subsurface layer,
the purple field lines
extend to both the main and parasitic flux tubes.
The current sheet is developed
between the two flux systems
immediately above the sheared PIL
in the photosphere.
Reflecting the smaller scale
of the PIL
(Figure \ref{fig:pil_test062}),
the purple flux is compact and very low-lying
compared to the other cases.
Contrary to Spot-Spot,
this purple flux
is located at the edge of the AR,
and is not trapped
by the overlying fields,
i.e., exposed to the outer space.

The Quadrupole and Inter-AR cases
somewhat resemble each other.
As explained in Section \ref{sec:general},
the two emerging flux systems
of the Quadrupole case
(yellow and green)
originate from the common single flux tube
and thus are connected beneath the surface.
Consequently,
the two photospheric polarities
show convergence motion
and become tightly packed.
Driven by this photospheric motion,
magnetic reconnection
between the two coronal loops occurs
in the current layer
with a sheet-like shape
extending in parallel with
the photospheric PIL.
Eventually, the purple flux system
is newly created,
which short-circuits the two inner polarities.

Although the two coronal loops
of the Inter-AR case
are not originally connected
beneath the surface
and, consequently, the contact
of the two flux systems
is less vigorous,
they in fact undergo magnetic reconnection
in the atmosphere
since the two bipoles expand
above the surface.
A vertically extending current sheet
is seen in between.
The two bipoles
eventually form
purple field lines
that connect the two independent ARs,
which may be related to
the flux rope
that erupted
as a fast CME
between NOAA 11944 and 11943
\citep{moe15}.

\subsection{Buildup of Magnetic Energy}\label{sec:energy}

In order to examine the accumulation
of magnetic energy
in the atmosphere,
we calculate the potential magnetic fields
from the $B_{z}$ map at $z_{\rm p}/H_{0}=2$
and measure the total magnetic energy
\begin{eqnarray}
  E_{\rm mag}=\int_{z\geq z_{\rm p}} \frac{\mbox{\boldmath $B$}^{2}}{8\pi}\, dV,
  \label{eq:emag}
\end{eqnarray}
potential energy
\begin{eqnarray}
  E_{\rm pot}=\int_{z\geq z_{\rm p}} \frac{\mbox{\boldmath $B$}_{\rm pot}^{2}}{8\pi}\, dV,
  \label{eq:epot}
\end{eqnarray}
and free energy
\begin{eqnarray}
  \Delta E_{\rm mag}\equiv E_{\rm mag}-E_{\rm pot}.
  \label{eq:efre}
\end{eqnarray}
Figure \ref{fig:evolution2} compares
the time evolutions of
$E_{\rm mag}$, $E_{\rm pot}$,
and $\Delta E_{\rm mag}$
for the four cases.
The time $\Delta t/\tau_{0}$
is measured since the flux appears
at the photosphere.
Here,
the simulation case with the largest energy
is the Spot-Spot case.
Reflecting the large photospheric flux
(Figure \ref{fig:evolution}),
it has a total energy and free energy
that are about one order of magnitude greater
than those of the other three cases.

The free energy of the Spot-Satellite case
is larger than
that of the Reference case
because free energy is stored
in the current layer
between the main and parasitic tubes
in the Spot-Satellite case
(see Figure \ref{fig:fl}, Spot-Satellite).
Since the current sheet lies
lower in the atmosphere,
where the density is higher
and the reconnection is less effective
(Section \ref{sec:setup:basic}),
the free energy is not significantly consumed
and gradually increases
over time.

The free energies of the remaining two cases
exhibit an interesting oscillatory behavior.
For example,
Quadrupole shows
a big bump around
$\Delta t/\tau_{0}=50$,
which corresponds to $t/\tau_{0}=270$.
This is because,
whereas the potential energy ($E_{\rm pot}$)
follows the monotonous growth
of the photospheric flux
(Figure \ref{fig:evolution}),
coronal reconnection
between the two magnetic loops
occurs when the two inner polarities
approach from $t/\tau_{0}=240$
(Figure \ref{fig:general_c})
and the actual magnetic energy ($E_{\rm mag}$)
starts reduction,
leading to the drastic loss
of the free energy.
The free energy decrease
of Inter-AR
after $\Delta t/\tau_{0}=40$
(corresponding to $t/\tau_{0}=260$)
is also due to the coronal reconnection
of the two magnetic systems,
which occurs later than in the Quadrupole case
because the two systems are significantly separated
from each other.

The bottom panel of Figure \ref{fig:evolution2}
is intended to compare the four cases
under the condition that
they have similar AR scales.
For each simulation case,
we normalize the free energy
by the three-halves power
of its total unsigned flux
at the final stage ($t/\tau_{0}=300$),
$\Phi_{\rm final}^{3/2}$.
Note that the AR area is approximately proportional
to the total unsigned flux, $\Phi$,
because the photospheric field is almost uniquely determined
by the pressure balance
between the magnetic field and the external gas.
Considering that the AR volume
is roughly proportional to the area to the $3/2$,
here we normalize the free magnetic energies
of the four different cases
with $\Phi^{3/2}$.

One can find that,
since the difference
among the four cases
in the bottom panel
is less prominent
than that in the middle panel,
the free energy depends highly
on the photospheric total flux
of each AR
(or equivalently area or volume).
However,
as seen from the bottom panel,
even for ARs
of the similar size scales,
the stored free energy may differ much,
up to a factor of five,
depending on the twist and geometrical configuration
of subsurface emerging fields.

\section{Flare and CME Predictions Based on Photospheric Measurements}\label{sec:sharp}

\subsection{Flare Predictions}

The prediction
of flares and CMEs
is currently one of the most important topics
of solar-terrestrial physics.
Since the measurement of the photospheric parameters
from vector magnetic data
is much easier
than the reconstruction of
full 3D magnetic fields,
most of the current flare prediction schemes
are based on such photospheric parameters.
After \cite{lek03} and \citet{bar07} made use of
the vector magnetogram
for flare prediction,
\citet{bob15} extracted various parameters
(including those suggested
by \citealt{lek03}, \citealt{sch07}, and \citealt{fis12})
from the {\it SDO}/HMI vector magnetogram
for each AR
\citep[SHARP data:][]{bob14}
and obtained good predictive performance
for flares of $\geq$ M1.0-class
using a machine learning algorithm.
By adding flare history and ultraviolet observables
to the SHARP parameters,
\citet{nis17} further developed
flare prediction models
with even higher performance
\citep[see also][]{mur15,liu17}.
The SHARP parameters
used by these authors are
summarized
in Table \ref{tab:sharp}.
The $F$-score (Fisher score) in the table indicates
the scoring of the parameter
given by \citet{bob15}.

Then,
the question is as follows.
Why do some of these parameters predict
the flare eruptions well
(indicated by larger $F$-scores),
while the others do not
(smaller $F$)?
The series of numerical simulations
of the present work,
which successfully reproduced
a variety of complex, non-potential configurations
of flare-productive ARs,
is one of the best ways to resolve
this mystery.
It is worth noting that
\citet{gue17} recently examined
various photospheric parameters
using simulations
proposed by \citet{lea13,lea14}
and found that parameters
related to the PIL
are the best to describe
the flare occurrence.
However,
their analysis was restricted
by the limited size of ARs
and complexity
of the simulations,
which could be compensated for
by our simulations.

Considering that the flare occurrence is
a releasing process
of free magnetic energy
stored in the atmosphere,
we compare the SHARP parameters
in Table \ref{tab:sharp}
and the stored free energy,
i.e., the maximum flare energy
that could potentially be released,
for the four simulation cases.
The top six panels
of Figure \ref{fig:sharp}
show samples of the comparisons.
In each diagram,
the horizontal axis
represents a SHARP parameter in question
that is measured at the surface
($z_{\rm p}/H_{0}=2$)
every $\Delta t/\tau_{0}=2$
after the flux appears
at the surface,
whereas the vertical axis represents
the stored free magnetic energy
at each moment
($\Delta E_{\rm mag}/E_{0}$,
measured directly
from the 3D computational domain:
see Section \ref{sec:energy}).
One can find that
some SHARP parameters
have strong proportionalities
with the free energy
(e.g., \textsc{totusjh} and \textsc{totpod}),
while others do not
(e.g., \textsc{epsy} and \textsc{epsx}).
It is reasonable that
parameters such as \textsc{totusjh}
(total unsigned current helicity)
show high correlations
because the free energy is stored
in the form of electric current.

For each diagram,
we compute
the correlation coefficient, $CC$,
in a log-log plot,
which indicates
how accurately
a given SHARP parameter reflects
the stored free energy,
and show it in the bottom right
of the diagram.
For each plot,
we assume each data point to be independent
and simply derive $CC$ from all data points,
regardless of the simulation cases.
Therefore,
there is only one $CC$
for each plot.
The $CC$ values for all 25 parameters are
summarized in Table \ref{tab:sharp}.

The bottom panel of Figure \ref{fig:sharp}
is a scatter plot
of absolute correlation, $|CC|$,
vs. $F$-score
for all SHARP parameters.
It is clearly seen that
parameters with larger $F$
yield larger $|CC|$.
In other words,
the SHARP parameters
that are excellent
in predicting the flare events
can predict the free energy
in the atmosphere
very accurately.
On the other hand,
the smaller-$F$ parameters
have weaker to almost no correlation
with the free energy,
indicating that
they are incapable of
predicting the free energy.
It should be noted that,
in each scatter plot,
the correlation is strong (weak)
because all four simulations
show consistently strong (weak) correlations.
For example,
$|CC|$ of \textsc{totpot} is 0.95,
which is due to high correlations
of the four cases:
0.89 (Spot-spot),
0.85 (Spot-satellite),
0.82 (Quadrupole),
and 0.98 (Inter-AR).

The above relationship
between $|CC|$ and $F$
confirms
the suggestion by \citet{wel09}
that parameters
strongly associated
with the flare activity
are extensive
(scaling with AR size:
indicated by ``E'' in Table \ref{tab:sharp})
because the free energy
is likely stored on large scales
and non-local.

The even higher prediction rates
obtained by \citet{nis17}
are probably because,
besides adding flare history,
they include
ultraviolet observables
that are sensitive to
chromospheric dynamics
such as triggering processes
(preflare brightenings)
before the flares occur
(see, e.g., \citealt{bam13}
for detailed observational analysis
of the flare triggers).
This suggests that,
although the SHARP parameters
properly indicate the accumulation
of free energy,
this is not sufficient to accurately predict
the exact occurrence of flares,
and we need
additional observables
that represent
the triggering of the flares.

\subsection{CME Predictions}

The rightmost column
of Table \ref{tab:sharp}
shows the ranking of SHARP parameters
for predicting CME eruptions,
as reported by \citet{bob16}.
Contrary to the flare prediction case,
the parameters
that successfully predict
whether a given flare produces a CME
are mostly intensive
(independent of AR size:
indicated by ``I'' in Table \ref{tab:sharp}),
not extensive.
One can also see from
this table that
they show
moderate to weak negative correlations
($CC<0$)
between the SHARP parameters
and the stored free energy.
This is because these parameters
are, in most cases,
normalized by the AR size
or some relevant factors.
That is,
the CME-predictive parameters
are not perfectly independent of the AR size;
rather, they have negative dependence
with the AR scale.

For example,
according to \citet{bob16},
\textsc{meangbt}
is one of the highest-ranking CME parameters,
and it is the sum of the horizontal magnetic gradient
normalized by the SHARP patch pixels $N$
(representing AR area):
\begin{eqnarray}
  \overline{|\nabla B_{\rm tot}|}
  =\frac{ \sum \sqrt{
    \left(\frac{\partial B}{\partial x}\right)^{2}
    +\left(\frac{\partial B}{\partial y}\right)^{2}
  }}{N}.
  \label{eq:meangbt}
\end{eqnarray}
Since the flare-causing PILs
tend to have a high magnetic gradient
(Section \ref{sec:intro}),
the numerator of Equation (\ref{eq:meangbt})
becomes larger for flaring ARs.
However,
the normalization by area
cancels this trend,
leading to
a negative correlation
with the free energy
($CC=-0.530$: Figure \ref{fig:sharp})
and thus a worse flare prediction rate
($F=192.3$).
Still,
this parameter yields
a high CME prediction performance.
This may also be
due to the normalization effect.
That is,
while the local flaring zone
is characterized by the field gradient
(numerator),
the global scale of the AR
is represented by the AR area
(denominator),
and the relationship between
the two factors
(their ratio)
determines the CME productivity.

The above discussion
is further supported by
observational studies.
\citet{sun15} investigated
the flare-rich but CME-poor AR NOAA 12192
and found that
this AR has a relatively weak non-potentiality
with a relatively strong overlying field
and, thus, the flux ropes
fail to erupt as CMEs.
They concluded that
CME eruption is described
by the relative measure of non-potentiality
over the restriction of the background field.
A statistical analysis by \citet{tor17}
revealed that
CME-less ARs show,
on average,
smaller $S_{\rm ribbon}/S_{\rm spot}$
(flare ribbon area normalized by total sunspot area)
and $|\Phi|_{\rm ribbon}/|\Phi|_{\rm AR}$
(total unsigned magnetic flux in the ribbon
normalized by the total unsigned flux
of the entire AR).
They argued that
the magnetic relation
between the large-scale structure of an AR
and the localized flaring domain
within it
is a key factor
determining the CME eruption.
We shall leave the detailed numerical investigation
on the relation
for future work.

\section{Summary}\label{sec:summary}

In this paper,
aiming at understanding
the creation
of flare-productive ARs,
especially the formation processes
of $\delta$-spots and sheared PILs
and the accumulation of free magnetic energy,
we performed flux emergence simulations
of four typical types
of flaring ARs
\citep{zir87,tor17}.
The four simulations share
similar numerical conditions
(tube's initial axial field,
background atmosphere, etc.)
in order that one can easily
compare the results.
The main results of this study
are summarized as follows.

The first category
of the four types of ARs,
Spot-Spot,
is modeled by an emergence
of a tightly-twisted, kink-unstable flux tube
from the convection zone
\citep{fan99,tak15}.
Because of the kink instability,
the tube's ascent is fastest
among the four cases.
The flux tube eventually produces
a complex AR
composed of two main sunspots
of both polarities
with extended tails,
which is probably classified
as $\beta\gamma\delta$.
As the main spots develop,
an elongated sheared PIL
spanning over the entire AR
is created at the center,
which shows a stripe pattern
of both polarities
that is highly reminiscent
of the magnetic channel
\citep{zir93}.
Owing to magnetic reconnection
of the two loop systems,
sheared arcade fields
are newly formed
above the central PIL.
Although this configuration has no clear access
to the outer atmosphere,
the quadrupolar structure achieved here
(main spots and extended tails)
is preferable for CME
\citep[e.g.][]{ant99,hir01}.
This AR possesses
the largest unsigned flux
in the photosphere
with the largest free energy
in the corona.
That is,
even if we use the flux tubes
with the same axial flux,
the photospheric unsigned flux can differ
by up to one order of magnitude
depending on the initial twist and geometry.

The Spot-Satellite AR
is achieved by the interaction
of main and parasitic flux tubes.
As the parasitic tube appears
at the photosphere
in the close vicinity
of the main spot,
they form tiny $\delta$-spots
at the edge of the AR
with a compact sheared PIL
($\beta\gamma\delta$ spot).
The newly formed field lines,
created through magnetic reconnection
between the two flux tubes,
are clearly exposed
to the upper atmosphere.

The Quadrupole AR
is modeled by the emergence
of a single flux tube
that rises at two buoyant sections
\citep{tor14,fang15}.
The two emerging bipoles
collide against each other
at the center
and create
a strongly packed $\delta$-spot
and a sheared PIL
with the highest $B_{z}$ gradient
(classified as $\beta\gamma$ or $\beta\gamma\delta$).
The coronal free energy
shows a fluctuation with time,
coincident with the magnetic reconnection
of the two emerging bipoles.

The two flux tubes
of the Inter-AR case,
placed in the convection zone
in a parallel fashion,
totally separated from each other,
produce two independent ARs
on the solar surface.
The two tubes, however, undergo
magnetic reconnection
in the corona
as they expand
in the atmosphere,
and consequently,
a new flux system
connecting the two ARs
is formed.
Although they have no clear sheared PIL
nor a $\delta$-spot
and thus should be classified
as two simple $\beta$-spots,
they in fact can produce X-class flares
if the free energy is sufficiently accumulated,
which could launch a fast CME
\citep{moe15,wan15}\footnote{Like other simulation cases, the CME-predictive SHARP parameters of the Inter-AR case show inversely correlated trends with the free energy (see e.g. \textsc{meangbt} of Figure \ref{fig:sharp}), which supports the possibility of CME eruptions from the Inter-AR configuration.}.

The sheared PIL
in a $\delta$-spot
is formed by the combination
of different factors
(Figure \ref{fig:inductioneq}).
The enhancement of the sheared field
is at first due to the advection.
As the spots of opposite polarities
come closer to each other,
they transport the horizontal flux
to the PIL in between.
Then,
the (relative) drifting motion
of the two spots
stretches the horizontal flux,
and thus, the shear component grows further.
The horizontal compression,
which is the pressing motion
caused by the two approaching spots,
also contributes to
the intensification
of the horizontal flux.

Some of
the SHARP parameters,
the photospheric observables
obtained from vector magnetograms,
are known to predict the solar flares
with high accuracy
\citep{bob15}.
From the $\delta$-spot models
of this paper,
we confirmed that
these parameters reflect
the free magnetic energy
stored in the corona
very well.
Since the free energy
is a global (non-local) value,
the extensive parameters,
i.e., those scaling with the AR size,
show higher prediction scores
\citep{wel09}.
For even better flare forecast,
we may need to add parameters
that are sensitive to
the triggering of the flares,
such as chromospheric brightenings
\citep{nis17,mur15}.

On the other hand,
it was also found that
most of the CME-predictive SHARP parameters
\citep{bob16}
do not reflect the coronal free energy well:
they show moderate to lower correlation
with the free energy.
These parameters are the values
normalized by the AR size
or some relevant factors.
This indicates the importance
of the magnetic relation between local flaring zones
(e.g. erupting flux rope)
and large-scale circumstances
(e.g. overlying arcades)
for CME productivity
\citep{sun15,tor17}.

\section{Discussion}\label{sec:discussion}

From the numerical simulations,
we derived the close connections
among the subsurface history of emerging flux,
the free magnetic energy
stored in the atmosphere,
and the various SHARP parameters
measured at the surface.
For example,
the flare-predictive parameters
are strongly correlated with the free energy
(Figure \ref{fig:sharp}),
while the free energy
highly depends
on the types of the emerging flux
(Figure \ref{fig:evolution2}).
These relationships obtained in this work
provide important information
such as which parameters are suitable,
not only for predicting flares and CMEs,
but also for probing
the subsurface state of emerging flux
that builds up flare-productive ARs.

Although we revealed the detailed formation
of $\delta$-spots and sheared magnetic structures,
how the complexity
of subsurface emerging flux
is produced
remained unclear.
Recently,
\citet{jae16} found that,
whereas the fractions of
all $\alpha$- and $\beta$-sunspots
remain constant over solar cycles
(roughly 20\% and 80\%, respectively),
the fraction of complex ARs,
appended with $\gamma$
and/or $\delta$,
increases drastically
from less than 10\%
at solar minimum
to more than 30\% at maximum.
From this result,
they suggested the possibility that
complex ARs are produced
by the collision
of simple ARs
around the surface layer
due to the higher frequency
of flux emergence
during solar maximum.

This situation
is more in favor of
the Spot-Satellite model,
in which two flux systems
interact with each other
in the subsurface region
\citep{fan98}.
Such interaction
may be a stochastic process,
probably coupled with convective dynamics.
Therefore,
we need global dynamo simulations
to investigate
how the emerging fluxes
interact with each other
and with convection cells
\citep[e.g.][]{nel14}.

Another remaining problem
is that we did not observe
any flare eruptions
in the simulations.
To follow the full story
from the emergence to eruption
including free energy accumulation,
current sheet formation,
and reconnection onset
\citep{man04,arc14,lea14},
we may need to improve the model,
for example,
by tracing an even longer evolution
in a wider simulation domain
(see \citealt{oi17} for emergence-to-eruption simulation
of the Quadrupole case).
Photospheric and subsurface convection
may supply magnetic shear and affect
long-term evolution of magnetic configuration,
which should be investigated further
in future
\citep{fan12,cha16}.

\acknowledgments

The authors are grateful
to the anonymous referee
for improving the manuscript.
S.T. and S.T. would like to thank
M.C.M. Cheung
for providing the potential field calculation code.
This work was supported by
JSPS KAKENHI Grant Numbers JP16K17671, JP15H05814,
JP16J02063.
Numerical computations were carried out on Cray XC30
at Center for Computational Astrophysics,
National Astronomical Observatory of Japan.

%

\vspace{5mm}





\appendix

\section{Visualization of Current Sheets}\label{sec:current}

\begin{figure*}
  \begin{center}
    \includegraphics[width=170mm]{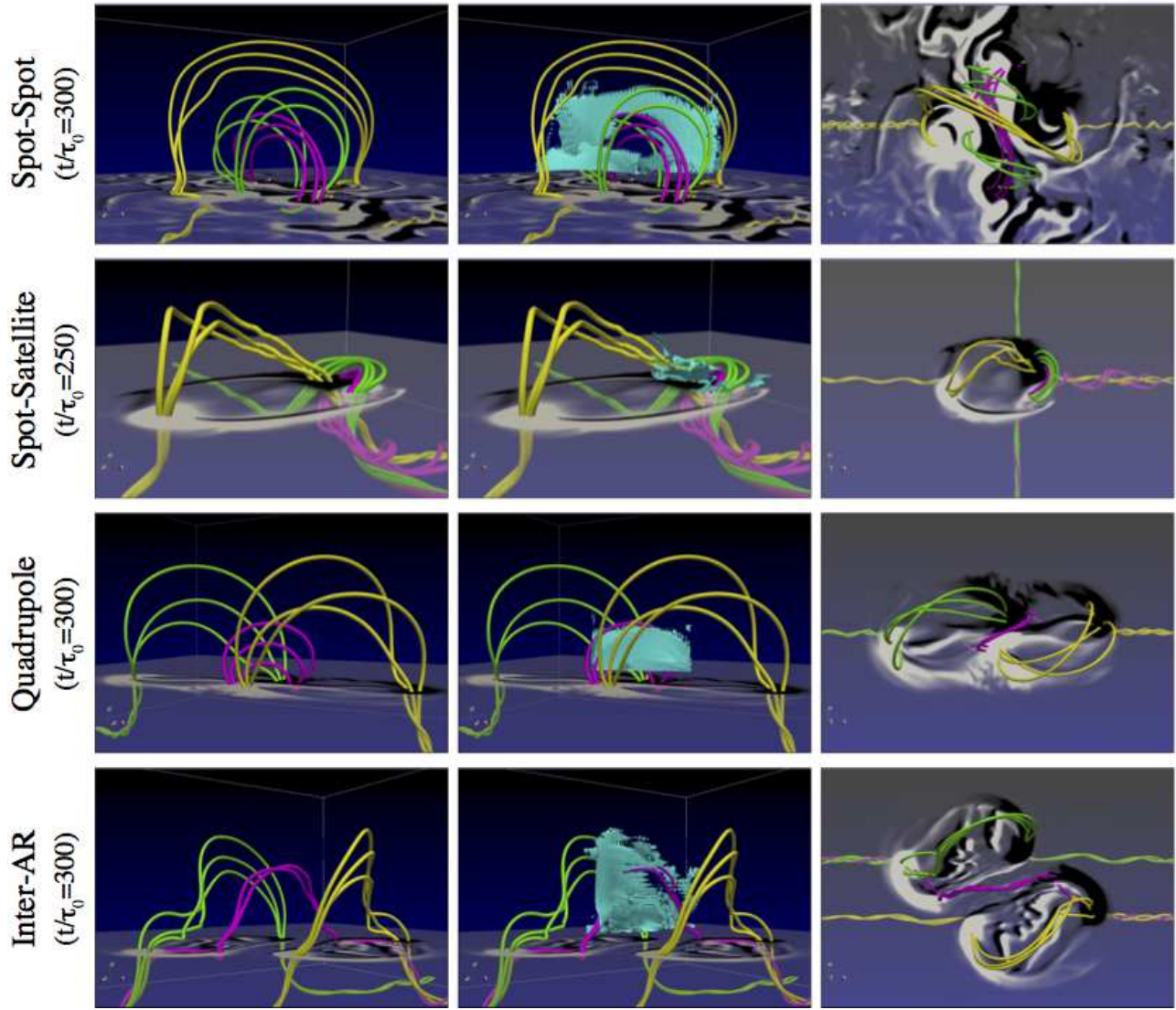}
  \end{center}
  \caption{3D magnetic structures for the four simulation cases.
    Surface magnetogram
    saturates at $B_{z}/B_{0}=\pm 0.3$,
    with reduced transparency
    for weaker field regions.
    See main text for explanations
    of the colors of the field lines.
    In the middle column,
    the electric current sheets
    are overplotted with sky blue isocontours
    (see Appendix \ref{sec:current}
    for the definition of the current sheets).
    \label{fig:fl}}
\end{figure*}

\begin{figure*}
  \begin{center}
    \includegraphics[width=80mm]{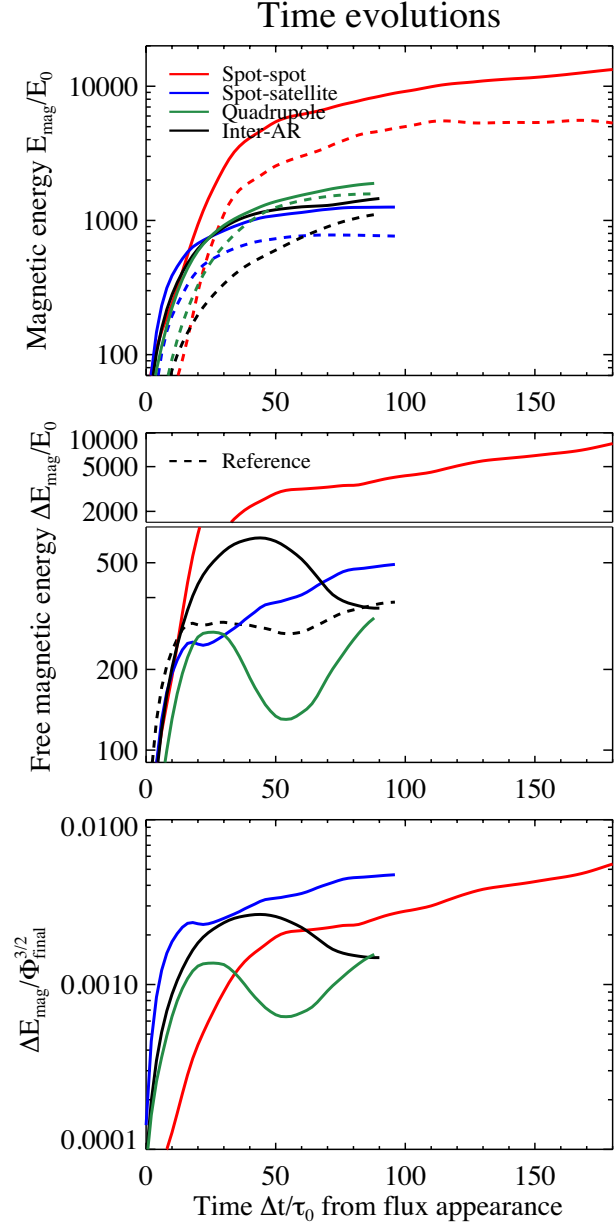}
  \end{center}
  \caption{(Top) Time evolution
    of the magnetic energy
    in the atmosphere
    for the four cases.
    Time $\Delta t/\tau_{0}$ is measured
    since the flux appears at the surface
    (see Figure \ref{fig:evolution}).
    Solid line indicates
    the actual total magnetic energy,
    $E_{\rm mag}$ (Equation (\ref{eq:emag})),
    while the dashed line is
    the calculated potential energy,
    $E_{\rm pot}$ (Equation (\ref{eq:epot})).
    (Middle) Evolution
    of the free magnetic energy,
    $\Delta E_{\rm mag}\equiv E_{\rm mag}-E_{\rm pot}$.
    The Reference case
    is also plotted
    with a dashed line.
    (Bottom) Free energy normalized
    by the three-halves power
    of its final ($t/\tau_{0}=300$)
    photospheric unsigned magnetic flux,
    $\Phi_{\rm final}^{3/2}$.
    \label{fig:evolution2}}
\end{figure*}

\begin{figure*}
  \begin{center}
    \includegraphics[width=80mm]{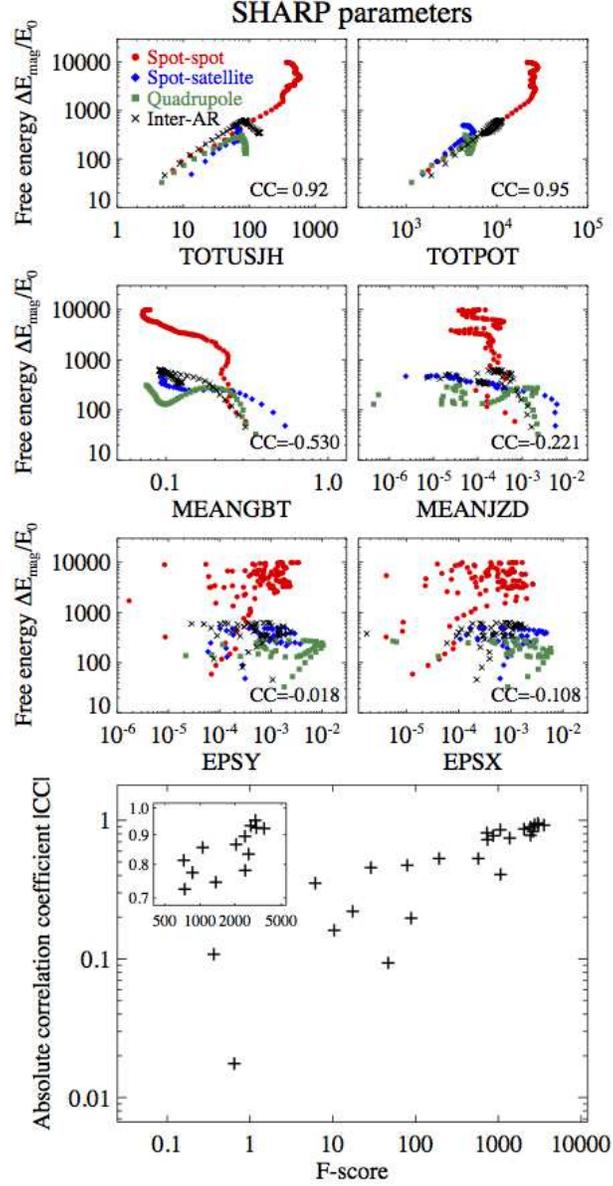}
  \end{center}
  \caption{(Top) Six sample diagrams showing
    free magnetic energy $\Delta E_{\rm mag}/E_{0}$
    vs. SHARP parameters,
    which are
    \textsc{totusjh}
    (total unsigned current helicity),
    \textsc{totpot}
    (total photospheric magnetic free energy density),
    \textsc{meangbt}
    (mean gradient of total field),
    \textsc{meanjzd}
    (mean vertical current density),
    \textsc{epsy}
    (sum of $y$-component of normalized Lorentz force),
    and \textsc{epsx}
    (sum of $x$-component of normalized Lorentz force):
    see Table \ref{tab:sharp} for detailed formulae.
    For the four simulations,
    the SHARP parameters are measured
    in the horizontal plane
    at $z_{\rm p}/H_{0}=2$
    with time steps of $\Delta t/\tau_{0}=2$
    after the flux appears
    at $z_{\rm p}/H_{0}=2$.
    Correlation coefficient, $CC$,
    calculated on the log-log plot
    is shown at the bottom right
    of each diagram.
    (Bottom) Scatter plot of
    absolute $CC$
    for all 25 SHARP parameters
    vs.
    $F$-score
    given by \citet{bob15},
    which indicates how well
    a given parameter predicts
    flare events.
    \label{fig:sharp}}
\end{figure*}

From Equation (\ref{eq:current}),
we can roughly estimate the thickness of the current sheet,
$\delta$:
\begin{eqnarray}
  J\sim \frac{1}{4\pi}
  |\mbox{\boldmath $\nabla$}\times\mbox{\boldmath $B$}|
  \sim
  \frac{1}{4\pi}
  \frac{B}{\delta},
\end{eqnarray}
or
\begin{eqnarray}
  \delta\sim
  \frac{1}{4\pi}
  \frac{B}{J}.
\end{eqnarray}
In the numerical simulations,
the magnetic field lines start reconnection
when the current thickness becomes
comparable to the grid spacing.
If we express the typical grid size
as $\Delta=\min{(\Delta x, \Delta y, \Delta z)}$,
the non-dimensional parameter
\begin{eqnarray}
  \widehat{J}
  =\frac{\Delta}{\delta}
  =\frac{4\pi J}{B}\Delta
\end{eqnarray}
approaches unity
in the core of
the current layer.
In Figure \ref{fig:fl},
we show the region
of $\widehat{J}\geq 8\times 10^{-3}$
with isocontours (sky blue)
instead of simply plotting $J$.

\begin{deluxetable}{llccccc}
\tabletypesize{\scriptsize}
\tablecaption{Properties of Flare Events\label{tab:sharp}}
\tablecolumns{7}
\tablewidth{0pt}
\tablehead{
\colhead{Keyword} & \colhead{Description} & \colhead{Formula} &
\colhead{$F$-Score} & \colhead{$CC$} & \colhead{Scaling} &
\colhead{CME Rank}
}
\decimals
\startdata
\textsc{totusjh} &
Total unsigned current helicity &
$H_{c_{\rm total}}\propto \sum |B_{z}\cdot J_{z}|$ &
3560 & $ 0.922$ & E & 16\\
\textsc{totbsq} &
Total magnitude of Lorentz force &
$F\propto \sum B^{2}$ &
3051 & $ 0.925$ & E & \nodata\\
\textsc{totpot} &
Total photospheric magnetic free energy density &
$\rho_{\rm tot}\propto \sum (\mbox{\boldmath $B$}^{\rm Obs}-\mbox{\boldmath $B$}^{\rm Pot})^{2}dA$ &
2996 & $ 0.952$ & E & 8\\
\textsc{totusjz} &
Total unsigned vertical current &
$J_{z_{\rm total}}=\sum |J_{z}|dA$ &
2733 & $ 0.933$ & E & 12\\
\textsc{absnjzh} &
Absolute value of the net current helicity &
$H_{c_{\rm abs}}\propto \left| \sum B_{z}\cdot J_{z}\right|$ &
2618 & $ 0.833$ & E & 13\\
\textsc{savncpp} &
Sum of the modulus of the net current per polarity &
$J_{z_{\rm sum}}\propto \left|\sum^{B_{z}^{+}} J_{z}dA\right|+\left|\sum^{B_{z}^{-}} J_{z}dA\right|$ &
2448 & $ 0.781$ & E & 18\\
\textsc{usflux} &
Total unsigned flux &
$\Phi=\sum |B_{z}|dA$ &
2437 & $ 0.894$ & E & 10\\
\textsc{area\_acr} &
Area of strong field pixels in the active region &
${\rm Area}=\sum {\rm Pixels}$ &
2047 & $ 0.865$ & E & 14\\
\textsc{totfz} &
Sum of $z$-component of Lorentz force &
$F_{z}\propto \sum(B_{x}^{2}+B_{y}^{2}-B_{z}^{2})dA$ &
1371 & $ 0.745$ & E & \nodata\\
\textsc{meanpot} &
Mean photospheric magnetic free energy &
$\overline{\rho}\propto \frac{1}{N}\sum (\mbox{\boldmath $B$}^{\rm Obs}-\mbox{\boldmath $B$}^{\rm Pot})^{2}$ &
1064 & $-0.406$ & I & 5\\
\textsc{r\_value} &
Sum of flux near polarity inversion line &
$\Phi=\sum|B_{LoS}|dA$ within $R$ mask &
1057 & $ 0.855$ & E & 15\\
\textsc{epsz} &
Sum of $z$-component of normalized Lorentz force &
$\delta F_{z}\propto \frac{\sum(B_{x}^{2}+B_{y}^{2}-B_{z}^{2})}{\sum B^{2}}$ &
864.1 & $-0.774$ & I & \nodata\\
\textsc{shrgt45} &
Fraction of Area with shear $> 45^{\circ}$ &
Area with shear $> 45^{\circ}$ / total area &
740.8 & $ 0.725$ & I & 7\\
\textsc{meanshr} &
Mean shear angle &
$\overline{\Gamma}=\frac{1}{N} \sum\arccos{\left( \frac{\mbox{\boldmath $B$}^{\rm Obs}\cdot\mbox{\boldmath $B$}^{\rm Pot}}{|B^{\rm Obs}||B^{\rm Pot}|} \right)}$ &
727.9 & $ 0.813$ & I & 6\\
\textsc{meangam} &
Mean angle of field from radial &
$\overline{\gamma}=\frac{1}{N}\sum \arctan{\left( \frac{B_{h}}{B_{z}}\right)}$ &
573.3 & $-0.535$ & I & 11\\
\textsc{meangbt} &
Mean gradient of total field &
$\overline{|\nabla B_{\rm tot}|}=\frac{1}{N}\sum \sqrt{\left(\frac{\partial B}{\partial x}\right)^{2}+\left(\frac{\partial B}{\partial y}\right)^{2}}$ &
192.3 & $-0.530$ & I & 4\\
\textsc{meangbz} &
Mean gradient of vertical field &
$\overline{|\nabla B_{z}|}=\frac{1}{N}\sum \sqrt{\left(\frac{\partial B_{z}}{\partial x}\right)^{2}+\left(\frac{\partial B_{z}}{\partial y}\right)^{2}}$ &
88.40 & $-0.197$ & I & 19\\
\textsc{meangbh} &
Mean gradient of horizontal field &
$\overline{|\nabla B_{h}|}=\frac{1}{N}\sum \sqrt{\left(\frac{\partial B_{h}}{\partial x}\right)^{2}+\left(\frac{\partial B_{h}}{\partial y}\right)^{2}}$ &
79.40 & $-0.474$ & I & 1\\
\textsc{meanjzh} &
Mean current helicity ($B_{z}$ contribution) &
$\overline{H_{c}}\propto \frac{1}{N}\sum B_{z}\cdot J_{z}$ &
46.73 & $-0.094$ & I & 2\\
\textsc{totfy} &
Sum of $y$-component of Lorentz force &
$F_{y}\propto \sum B_{y}B_{z}dA$ &
28.92 & $ 0.456$ & E & \nodata\\
\textsc{meanjzd} &
Mean vertical current density &
$\overline{J_{z}}\propto \frac{1}{N}\sum \left(\frac{\partial B_{y}}{\partial x}-\frac{\partial B_{x}}{\partial y}\right)$ &
17.44 & $-0.221$ & I & 9\\
\textsc{meanalp} &
Mean characteristic twist parameter, $\alpha$ &
$\alpha_{\rm total}\propto \frac{\sum J_{z}\cdot B_{z}}{\sum B_{z}^{2}}$ &
10.41 & $-0.161$ & I & 3\\
\textsc{totfx} &
Sum of $x$-component of Lorentz force &
$F_{x}\propto -\sum B_{x}B_{z}dA$ &
6.147 & $ 0.352$ & E & \nodata\\
\textsc{epsy} &
Sum of $y$-component of normalized Lorentz force &
$\delta F_{y}\propto\frac{-\sum B_{y}B_{z}}{\sum B^{2}}$ &
0.647 & $-0.018$ & I & \nodata\\
\textsc{epsx} &
Sum of $x$-component of normalized Lorentz force &
$\delta F_{x}\propto\frac{\sum B_{x}B_{z}}{\sum B^{2}}$ &
0.366 & $-0.108$ & I & \nodata\\
\enddata
\tablecomments{For descriptions and formulae
  of the SHARP parameters,
  we follow the original notations
  of \citet{bob15}.
  In the analysis of the simulation results
  of this paper,
  we measured each parameter
  at the $z_{\rm p}/H_{0}=2$ plane
  every $\Delta t/\tau_{0}=2$
  after magnetic flux appears
  at $z_{\rm p}/H_{0}=2$.
  The threshold for the absolute field strength
  above which the total and mean values are calculated
  is selected to be $B/B_{0}=0.04$ (equivalently 10 G),
  while the threshold for the ``strong field''
  (\textsc{area\_acr})
  and for measuring
  \citet{sch07}'s $R$
  (\textsc{r\_value})
  is $B/B_{0}=0.2$ (500 G).
  $F$-score here is
  the scoring of the parameters
  for predicting solar flares
  provided by \citet{bob15},
  while $CC$ is the correlation coefficient
  obtained from the four simulation cases
  by comparing the free magnetic energy
  and SHARP parameter
  at each moment
  (see top panels of Figure \ref{fig:sharp}).
  Following \citet{wel09},
  we classified the parameters
  into extensive (E),
  where a given parameter increases with AR size,
  and intensive (I),
  where the parameter
  is independent of AR size.
  The rightmost column shows
  the ranking of features
  for predicting CME eruptions,
  as reported
  by \citet{bob16}.}
\end{deluxetable}

\end{document}